%% file: main.tex
\documentclass[manuscript, review=False, screen]{acmart}
\usepackage{booktabs} % For formal tables

\usepackage[ruled]{algorithm2e} % For algorithms

\setcopyright{rightsretained}
\acmJournal{TOMPECS}
\acmYear{2020} \acmVolume{1} \acmNumber{1} \acmArticle{1} \acmMonth{1} \acmPrice{}\acmDOI{10.1145/3381996}

\usepackage[english]{babel}

\usepackage{import}
\usepackage[]{units}
\usepackage{nameref}
\usepackage{url}

\usepackage{hyperref}
\usepackage{cleveref}

\usepackage{subfigure}
\usepackage{tabularx}
\usepackage{multirow}

\usepackage{graphicx}
\usepackage[color]{changebar}
\usepackage{changes}

\cbcolor{orange}
\usepackage{mathtools}

\usepackage[inline]{enumitem}

\usepackage[np]{numprint}
\npstyleenglish

\usepackage{amssymb}
\usepackage{xcolor}

\usepackage{tikz}

\input{commands}
\begin{document}
	
% Activate this to add an IEEE at the top for the author version of the paper
% \copyrightstatement

\input{header}

% Abstract
\import{content/}{01_abstract}

 \begin{CCSXML}
<ccs2012>
<concept>
<concept_id>10003033.10003034.10003035</concept_id>
<concept_desc>Networks~Network design principles</concept_desc>
<concept_significance>300</concept_significance>
</concept>
<concept>
<concept_id>10003120.10003121</concept_id>
<concept_desc>Human-centered computing~Human computer interaction (HCI)</concept_desc>
<concept_significance>300</concept_significance>
</concept>
<concept>
<concept_id>10003752.10003809.10003716.10011138.10010041</concept_id>
<concept_desc>Theory of computation~Linear programming</concept_desc>
<concept_significance>300</concept_significance>
</concept>
<concept>
<concept_id>10010405.10010406.10003228.10003229</concept_id>
<concept_desc>Applied computing~Intranets</concept_desc>
<concept_significance>300</concept_significance>
</concept>
</ccs2012>
\end{CCSXML}

\ccsdesc[300]{Networks~Network design principles}
\ccsdesc[300]{Human-centered computing~Human computer interaction (HCI)}
\ccsdesc[300]{Theory of computation~Linear programming}
\ccsdesc[300]{Applied computing~Intranets}


\keywords{SDN, QoE, QoS, pacing, enterprise, network, HAS, VoIP, browsing}

\maketitle

\import{content/}{02_introduction}

\import{content/}{03_background}

\import{content/}{04_overview_architecture}

\import{content/}{05_application_utility}

\import{content/}{06_networkwide_app_fairness}

\import{content/}{07_eval_methodology}

\import{content/}{08_results}

%\import{content/}{09_discussion}

\import{content/}{10_conclusion}

\bibliographystyle{acm}
\bibliography{bibliography}

\newpage
\import{content/}{11_appendix}

%\newpage
%\tableofcontents

\end{document}

%% file: commands.tex
\newcommand*\circled[1]{\tikz[baseline=(char.base)]{
            \node[shape=circle,draw,inner sep=1pt] (char) {#1};}}
            
\newcommand*\rectangled[1]{\tikz[baseline=(char.base)]{
            \node[shape=rectangle,draw,inner sep=1.5pt] (char) {#1};}}

\newcommand{\fkey}[1]{(\circled{\footnotesize #1})}
\newcommand{\fkeynb}[1]{\circled{\footnotesize #1}}

\newcommand{\mbps}[1]{\unit[#1]{Mbps}}
\newcommand{\kbps}[1]{\unit[#1]{Kbps}}
\newcommand{\ms}[1]{\unit[#1]{ms}}
\newcommand{\perc}[1]{\unit[#1]{\%}}

\newcommand{\reftab}[1]{Table~\ref{#1}}

\newcommand{\vod}{\textit{VoD}}
\newcommand{\live}{\textit{Live}}
\newcommand{\voip}{\textit{VoIP}}
\newcommand{\web}{\textit{WEB}}
\newcommand{\dl}{\textit{DL}}
\newcommand{\ssh}{\textit{SSH}}

\newcommand{\managed}{$\Box$}
\newcommand{\besteffort}{$\circ$}
\newcommand{\target}{$\times$}

\newcommand{\mvod}{\text{\LARGE$\lhd$}}

\newcommand{\mvoip}{$\circ$}
\newcommand{\mweb}{\text{\LARGE$\diamond$}}
\newcommand{\mdl}{$\bigcirc$}
\newcommand{\mssh}{\text{\LARGE$\bigtriangleup$}}

\newcommand{\pacer}{\rectangled{P}} 

\newcommand{\vodwm}{\vod{}~(\mvod{})}

\newcommand{\voipwm}{\voip{}~(\mvoip{})}
\newcommand{\webwm}{\web{}~(\mweb{})}
\newcommand{\dlwm}{\dl{}~(\mdl{})}
\newcommand{\sshwm}{\ssh{}~(\mssh{})}

\let\OldReplaced\replaced
\renewcommand{\replaced}[2]{\cbstart\OldReplaced{#1}{#2}\cbend}

\let\OldAdded\added
\renewcommand{\added}[1]{\cbstart\OldAdded{#1}\cbend~}

\let\OldDeleted\deleted
\renewcommand{\deleted}[1]{\cbstart\OldDeleted{#1}\cbend~}

\newcommand{\blue}[1]{#1}

\newcommand{\UKPI}{\blue{M}}
\newcommand{\MOS}{\blue{MOS}}
\newcommand{\UV}{\blue{\theta}}
\newcommand{\UVMIN}{\blue{\UV^{(\text{min})}}}
\newcommand{\UVMINI}{\blue{\UV^{(\text{min,1})}}}
\newcommand{\UVMINII}{\blue{\UV^{(\text{min,2})}}}
\newcommand{\Ux}{\blue{U}}
\newcommand{\TU}{\blue{\Lambda}}
\newcommand{\UTP}{\blue{\tau}}
\newcommand{\UD}{\blue{\delta}}
\newcommand{\BTP}{\blue{\mathrm{T}}}
\newcommand{\BD}{\blue{\Delta}}
\newcommand{\DCU}{\blue{\psi}}
\newcommand{\DCD}{\blue{\Psi}}
\newcommand{\LU}{\blue{\Omega}}
\newcommand{\LD}{\blue{\omega}}
\newcommand{\AD}{\blue{\Upsilon}}
\newcommand{\TP}{\blue{\eta}}
\newcommand{\Dx}{\blue{D}}
\newcommand{\Cx}{\blue{C}}
\newcommand{\Fx}{\blue{F}}

\newcommand{\Sx}{\blue{S}}
\newcommand{\Tx}{\blue{T}}
\newcommand{\Ax}{\blue{\mathcal{A}}}

%% file: header.tex
\title{End-host Based Application Pacing for QoE-Fairness in Enterprise Networks}
\title{Application-aware Resource Allocation in Enterprise Networks With End-host Pacing}
\title{Predictable Application and User-aware Resource Allocation in\\Enterprise Networks With End-host Pacing}
\title{Dependable Application- and User-aware Resource Allocation in\\Enterprise Networks With End-host Pacing}
\title{Dependable Application- and User-aware Resource Allocation\\With End-host Pacing in Enterprise Networks}
\title{Dependable Application- and User-aware Resource Allocation\\in Enterprise Networks Using End-host Pacing}
\title{Scalable Application- and User-aware Resource Allocation\\in Enterprise Networks Using End-Host Pacing}
%\title{Dependable Application-Aware Resource Allocation\\in Enterprise Networks Using End-host Pacing}
% \titlenote{}
% \subtitle{}
% \subtitlenote{}

% Susanna:
%Using end-host pacing for reliable application- and user-aware resource allocation in enterprise networks. 
%Reliable application- and user-aware resource allocation in enterprise networks Using End-host pacing

\author{Christian Sieber}
\email{c.sieber@tum.de}
\orcid{0000-0003-3820-3674}
\affiliation{%
  \institution{Chair of Communication Networks, Technical University of Munich}
  \country{Germany}}

\author{Susanna Schwarzmann}
\email{susanna.schwarzmann@inet.tu-berlin.de}
\orcid{0000-0002-3705-7559}
\affiliation{%
 \institution{FG INET, TU Berlin}
 \country{Germany}}
 
\author{Andreas Blenk}
\email{andreas.blenk@tum.de}
\orcid{0000-0002-2001-4050}
\affiliation{%
  \institution{Chair of Communication Networks, Technical University of Munich}
  \country{Germany}
}
\affiliation{
	\institution{Faculty of Computer Science, University of Vienna}
	\country{Austria}
}

\author{Thomas Zinner}
\email{zinner@inet.tu-berlin.de}
\orcid{0000-0002-4179-4105}
\affiliation{%
 \institution{FG INET, TU Berlin}
 \country{Germany}}
  
\author{Wolfgang Kellerer}
\email{wolfgang.kellerer@tum.de}
\orcid{0000-0003-4358-8038}
\affiliation{%
  \institution{Chair of Communication Networks, Technical University of Munich}
  \country{Germany}}

\renewcommand\shortauthors{Sieber, C. et al.}
\renewcommand\shorttitle{Scalable Application- and User-aware Resource Allocation in Enterprise Networks}

%% file: content/01_abstract.tex
\begin{abstract}
\vspace{1em}

Providing scalable user- and application-aware resource allocation for heterogeneous applications sharing an enterprise network 
is still an unresolved problem.
The main challenges are: 
(i) How to define user- and application-aware shares of resources?
(ii) How to determine an allocation of shares of network resources to applications?
(iii) How to allocate the shares per application in heterogeneous networks at scale?
In this paper we propose solutions to the three challenges and introduce a system design for enterprise deployment.

Defining the necessary resource shares per application is hard, as the intended use case, the user's environment, e.g., big or small display, and the user's preferences influence the resource demand.
%For example web browser's and the HTTP protocol became the de-facto standard for many use cases such as video streaming, video conferencing and business applications.
We tackle the challenge by associating application flows with utility functions from subjective user experience models, selected Key Performance Indicators, and measurements.
The specific utility functions then enable a mapping of network resources in terms of throughput and latency budget to a common user-level utility scale.
A sensible distribution of the resources is determined by formulating a multi-objective mixed integer linear program to solve the throughput- and delay-aware embedding of each utility function in the network for a max-min fairness criteria.
The allocation of resources in traditional networks with policing and scheduling cannot distinguish large numbers of classes and interacts badly with congestion control algorithms.
We propose a resource allocation system design for enterprise networks based on Software-Defined Networking principles to achieve delay-constrained routing in the network and application pacing at the end-hosts.
%The pacing is dictated by the central controller and implemented through the agents on the end-hosts.

The system design is evaluated against best effort networks in a proof-of-concept set-up for scenarios with increasing number of parallel applications competing for the throughput of a constrained link.
The competing applications belong to the five application classes web browsing, file download, remote terminal work, video streaming, and Voice-over-IP.
The results show that the proposed methodology improves the minimum and total utility, minimizes packet loss and queuing delay at bottlenecks, establishes fairness in terms of utility between applications, and achieves predictable application performance at high link utilization.

\vspace{1em}
\end{abstract}

%% file: content/02_introduction.tex
\section{Introduction} \label{sec:introduction} 

Increasing bandwidth demands by multimedia-rich applications and low delay requirements for real-time communications present a challenge for modern enterprise network designs.
Despite a variety of demands, an enterprise network has to support the employees by providing a reliable infrastructure for the deployed network applications.
%The enterprise network has to provide employees with unobstructed access to network resources and IP communication services at all times.
%Every employee must have access to network resources as required by her/his work at any time.
Alongside the employees, the network resources are drained by automated processes such as backup transfers or by Internet of Things (IoT) devices such as surveillance cameras or sensors.
A network design is required which allocates every application its share of the available network resources while at the same time minimizes the need for over-provisioning.
%Over-provisioning of available resources is costly and, by explicit resource allocation, individual demands of applications or users can be considered, thus resulting in dependable application performance.
%
There are three main challenging research questions for application- and user-aware resource allocation in enterprise networks: 
\vspace{0.4em}
\begin{enumerate}[label=\Roman*)]
\itemsep0.4em 
\item \textit{Define}: How to define an application-aware allocation of resources in terms of Quality of Experience (QoE) of the user, considering the variety of application classes and their demands? 
\item \textit{Determine:} How to determine shares of resources for each application under resource constraints considering the definition of application-awareness derived in I)?
\item \textit{Allocate:} How to allocate each application its share of the network resources in heterogeneous enterprise networks where the availability of QoS mechanisms at each hop highly depends on the deployed switching hardware?
\end{enumerate}
\vspace{0.4em}

%\cbstart
Today there are commonly two high-level approaches for resource allocation in enterprise networks: best effort transport with sender-based congestion control and Quality of Service (QoS) mechanisms on the forwarding devices.
%Best effort networks implement resource allocation per flow at the end-hosts with sender congestion control.
%\cbend
But this is neither stable or fair in terms of goodput when applications compete for a link's bandwidth \cite{lukaseder2016comparison}, nor aware of the specific application or the user behind it.
This can lead to bad application quality and, as a consequence, to user dissatisfaction.
The second option, QoS configuration at the forwarding devices, either discards or delays data packets of an application or application class in favor of another class or application.
However, enforcing QoS on intermediate devices has several drawbacks. 
Buffer space and scheduling QoS options are limited on the devices. 
Discarding packets along the way from the sender to receiver interacts badly with the sender's congestion control~\cite{flach2016internet} and increases the network load due to the retransmission of discarded packets. 

Moving the QoS enforcement from the intermediate devices to the end-hosts is a viable third option, as shown by data-center operators: 
By using a central controller, network monitoring, and programmable application pacing at the sender and receiver, a specific amount of the available throughput can be allocated to each application.
Congestion in the network is then prevented by limiting the total sending rate of all applications \cite{kumar2015bwe}.
At the end-hosts, applications, i.e., the primary contributors to the network load, can be restricted from generating more data than the network can carry.
Furthermore, the limited QoS options, such as interface queues, can be reserved for high-profile use cases such as critical real-time traffic and separating managed from best-effort traffic.

%In this paper we explore the adaptation of this option for enterprise environments.
In this paper, we apply this concept to enterprise networks and show that, indeed, a global control strategy with end-host pacing can significantly improve user experience.
%To achieve this, we undertake three main steps, addressing the research questions listed above:  
%\fixme{The how is missing in the define-step}
%We I) \textit{define} application- and user-aware allocation for enterprise networks, II) formulate the problem of \textit{determining} the shares for each application based on an absolute utility scale tied to user experience and III) evaluate the \textit{allocation} through pacing for popular application classes.
Next we define the problems in detail.

\subsection{Problems Definitions}

In plain best effort networks, resource allocation is implemented on transport-level at the endpoints, e.g., at web servers and browsers, via TCP congestion control.
Congestion control works at sender-side by increasing or decreasing the sending rate based on observed packet loss and the Round-Trip Time (RTT).
TCP's goal is to divide the available data-rate equally between active TCP connections.
In the network, the data packets of a sending application, e.g., a web server, are treated equally by the forwarding devices.
If the receiving rate at a forwarding device's interface exceeds the maximum physical sending rate, packets are queued in a buffer or dropped if the buffer is full.

The main problems with plain best effort networks are:
\begin{enumerate*}
\item Some applications, such as web browsers, behave unfair and open multiple parallel TCP connections and therefore can receive a larger fraction of the available throughput.
\item Datagram-based applications, such as Voice-over-IP (VoIP), often do not implement any congestion control at all.
\item The effectiveness of TCP congestion control depends on factors such as the specific congestion control algorithm, delay, packet loss, relative start times of competing TCP flows and how active a TCP connection is.
\item Different demands of applications are not considered, e.g., in terms of minimum throughput and maximum delay. 
Thus, there is no application-awareness in best effort networks.
\end{enumerate*}

Commonly, the problems of best effort networks are addressed by enterprises by implementing QoS mechanisms in the network.
QoS mechanisms on the forwarding devices allow to prioritize some packets over others based on matching rules.
For example priority queuing allows to put VoIP packets based on the Type of Service (ToS) flag, VLAN tag or specific UDP ports into a queue with preferred treatment.
That way the delay and packet loss of VoIP calls is kept low and isolated from other traffic.
Flow- or class-based Weighted Fair Queueing (WFQ) allows to put individual application flows or whole application classes into separate queues with guaranteed minimum bandwidth.
Token bucket (TB) policing allows to limit the data-rate of individual flows or classes without the need for switch buffer space.
For example mobile service providers are known to use TB policing to limit the data-rate of video streaming services \cite{flach2016internet}.

But implementing QoS in the network is costly and inefficient:
\begin{enumerate*}
\item Buffers in forwarding devices are expensive and there is only a limited number of queues to configure per egress interface, typically about 8 \footnote{Jim Warner, \url{https://people.ucsc.edu/~warner/buffer.html}, last accessed: 11.10.2018}.
This is insufficient for implementing a sophisticated strategy to distinguish hundreds of active applications of multiple classes in a network.
\item Policing interacts badly with transport-level congestion avoidance algorithms resulting in lost packets.
Lost packets cause retransmissions and decrease transmission efficiency \cite{flach2016internet}.
\item Heterogeneous enterprise networks with diverse forwarding devices from different vendors are complex and error-prune to manage, hampering the enforcement of end-to-end QoS options.
Furthermore, there are no common QoS abstractions across switching hardware vendors.
Hence, deploying a single QoS strategy across devices might not be possible, especially if not all devices support the required features.
\item Encryption or header field ambiguity can prevent the correct identification of application classes in the network.
%\item There is no awareness of the experience of the user. 
%Applications are prioritized or downgraded based solely on the class of the application.
%\item There is no differentiation of applications of the same class.
%Hence all applications of a class are treated as equally important with the same demands.
%However, application classes do not identify the demand accurately.
%Classes do not consider specific needs for events, variations in demand, e.g., video streaming to smartphone vs. to large TV set, or different use cases of an application. 
%A user may watch a video, do video conferencing or read text-only articles in the browser.
 %, e.g. web browser for conferencing, video streaming or browsing.
%As a consequence applications are also unable to reliable determine the available goodput 
\end{enumerate*}

Hence, with limited or incompatible QoS mechanisms and the issues regarding identification of application flows, a scalable and application-aware network design is hard to implement in the network (see also Section \ref{sec:background}).

\subsection{Proposed Solutions}

We realize the resource allocation by implementing centrally-controlled pacing of individual applications at the end-hosts. 
Packet pacing at the end-hosts ensures that a stream of packets conforms to a specified data-rate by adding artificial delays between consecutive packets during the sending process.
Pacing prevents packet loss by smoothing out packet bursts and allows for shallow buffers in the intermediate forwarding nodes.
Shallow buffers reduce queuing delays and avoid expensive switch buffer space.
Applications can reliably determine their available goodput and it is unnecessary to probe the throughput by loss-based congestion control mechanisms.
Furthermore, pacing at the end-host allows for implementation of effective backpressure to the applications producing the data, reducing the amount of buffered data in the network stack.
Pacing at the end-hosts can scale to thousands of traffic classes \cite{saeed2017carousel}, congestion in the network can be avoided by a central management of the available resources \cite{kumar2015bwe} and application flows can be identified at the source.
Recent works show that bandwidth allocation to applications can be implemented hierarchically at global scale, enabling high percentages of link utilization \cite{kumar2015bwe}.
Sender congestion control and QoS in the network are downgraded both to failsafe solutions and supportive roles in the overall QoS strategy, e.g., in cases the central control fails or embedded devices cannot be modified.

%Pacing also reduces packet loss at high link utilizations and makes sender-based congestion avoidance unnecessary.

Ultimately, a user of an application does not care about what share of the resources is allocated to her/him as long as her/his user experience, or \textit{Quality of Experience} (QoE), with the application is positive.
For that reason, challenges I) and II), i.e., how to \textit{define} and \textit{determine} sensible allocations, are tackled based on the resulting user experience. 
%We introduce application- and user-awareness by a common utility scale between application classes and fine-grained utility functions for each application use case based on subjective user experience models.
%The central controller determines the shares and the delay-aware flow routing for each application based on the utility functions and allocates the resources through pacing at the end-hosts' network stacks.
We define the user experience as a per-application utility function of throughput and delay.
The utility function is derived from user experience models from the literature and selected application Key Performance Indicators (KPIs).
By jointly optimizing the utility and network resources usage, a fair share in terms of utility can be determined given a set of applications, utility functions, and constrained network resources.
Challenge III) is the scalable allocation of the calculated application shares.
We propose centrally-controlled application pacing at the end-hosts combined with per application flow routing.
Routing is solved implicitly by our problem formulation by selecting paths for application flows which satisfy capacity and delay requirements.
Routing per flow can then be implemented through Software-Defined Networking (SDN) for all applications in the network.
%SDN is the idea of splitting the data- and control plane of a network and moving all control decisions to a logically centralized entity, the network controller.
%SDN data-plane nodes are reduced to simple forwarding devices based on simple match-action rules pushed by the network controller.
The identification of application flows in the network, e.g., source and destination TCP/UDP ports, are provided to the central controller by software agents at the end-hosts.
%We extend the SDN principles via software-based agents to the edge of the network, to the end-hosts, e.g., desktop PCs, laptops or smartphones, and to the servers.
%That way, flows can by accurately identified by their originating source, e.g., a client application or a HTTP server, and enriched with KPIs provided by the application.
Applications can then be subjected to routing and pacing as dictated by the network controller.

\begin{figure}[t]
\centering
\includegraphics[width=390pt]{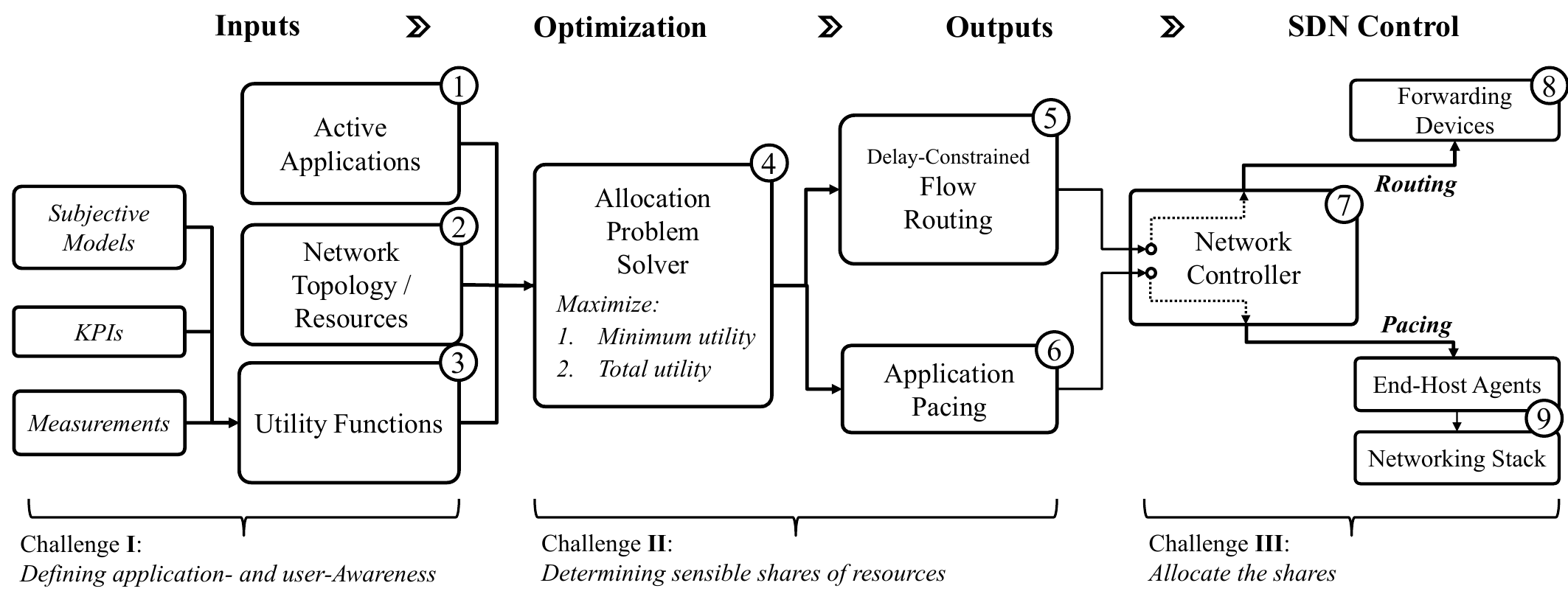}
\caption{Overview over the challenges and the proposed solution towards scalable application-aware resource allocation in enterprise networks.
Based on the set of applications \protect\fkey{1}, resources \protect\fkey{2} and application utility functions \protect\fkey{3}, the allocation problem solver maximizes the minimum and total utility over all active applications \protect\fkey{4}.
As a result, delay-constrained flow routing \protect\fkey{5} and application pacing rates \protect\fkey{6} are implemented by a network controller \protect\fkey{7} in the network \protect\fkey{8} and on the end-hosts \protect\fkey{9}.
}
\label{fig:introfig}
\Description[Three challenges: Define, Determine, Allocate]{Application and network metrics, an allocation problem solver and fine-grained network control provide application-aware resource allocation.}
\end{figure}

Figure~\ref{fig:introfig} summarizes the general methodology of the proposed solution.
First, the active applications \fkey{1} in the network are determined by end-host agents and network monitoring.
Second, the network topology and available resources \fkey{2} are known by the network controller.
Third, a suitable utility function \fkey{3} based on subjective QoE models, application KPIs, and measurements is associated with each application.
An allocation problem solver \fkey{4} then determines the per-application routing \fkey{5} and application pacing rates \fkey{6} based on a fairness criteria.
Routing rules are then implemented by the network controller \fkey{7} on the forwarding devices \fkey{8} and pacing rates are enforced at the end-hosts \fkey{9}.

The system is implemented as a proof-of-concept set-up with support for the following five application classes: web browsing, batch file transfer, VoIP, adaptive video streaming, and remote administration.
In the set-up, we evaluate static scenarios where a fixed number of parallel clients with multiple applications have to use a resource constrained link to communicate with central services, such as it is the case in SD-WAN or remote building scenarios.
The results show that central pacing can provide dependable application performance and increases inter-application fairness at high link utilizations.

\subsection{Contributions}

The contributions of this paper are as follows:
\begin{enumerate}
\item We present a system design for scalable user-aware resource allocation in enterprise networks based on SDN-principles and end-host pacing (Section \ref{sec:overview_architecture}).
The design does not make assumptions about the availability of QoS mechanisms such as WFQ on the forwarding devices.
%Furthermore, pacing at end-hosts can scale to large number of parallel applications.
%\item We select suitable user experience models from the literature for selected application classes and define utility functions based on the models.  \fixme{rewrite, we do not survey, we show utility functions}
\item We define throughput- and delay-dependent utility functions for five application classes. 
Furthermore, we discuss deployment options and trade-offs regarding the creation and accuracy of the utility functions (Section \ref{sec:applications}).
Compared to other works, the utility functions are based on actual subjective studies and thus tied to the experience of the user instead of technical KPIs.
Furthermore, measurements of the applications' behavior under limited available resources are used to determine the relationship between resources and user experience.
\item We formulate the utility throughput- and delay-aware allocation problem as a 2-step Mixed Integer Linear Program (MILP) with max-min fairness criteria.
The first step maximizes the minimum utility in the network (\textit{max-min-fairness}), while the second step maximizes the sum of all utilities for a constrained minimum utility (Section \ref{subsec:problem_formulation} and in detail in Appendix \ref{appendix:problem}).
While the min-max utility proportional fair bandwidth allocation problem is well studied in literature, the problem combination of bandwidth allocation and delay-aware routing for arbitrary utility functions is not formulated so far.
Note that in this paper we provide an optimal algorithm for the allocation problem, but with limited scalability.
\item We evaluate application mixes with over 100 parallel applications of 5 common use cases in a proof-of-concept set-up. 
The results show how pacing can improve delay and packet loss at bottlenecks and can significantly increase inter-application fairness in terms of utility.
Furthermore, pacing leads to predictable application performance even at high levels of network utilization (Section \ref{sec:results}).
\item We provide all material to the paper, such as the automated applications, a virtual experimentation set-up, and optimization formulation as open source software. \footnote{https://github.com/tum-lkn/appaware - Supplemental material to this article\label{fn:appaware}}
\end{enumerate}

\subsection{Paper Structure}

The paper is structured as follows.
Section \ref{sec:background} introduces the background and related work.
Section \ref{sec:overview_architecture} presents the proposed system architecture.
Afterwards we \textit{define} the shares (Section \ref{sec:applications}), \textit{determine} shares under resource constraints (Section \ref{subsec:problem_formulation}), discuss the \textit{allocation} of the shares in a experimental set-up (Section \ref{sec:evaluation}) and \textit{evaluate} the effectiveness of the proposed approach in the set-up (Section \ref{sec:results}).
Section \ref{sec:conclusion} summarizes the results, discusses future research directions and concludes this paper.

%% file: content/03_background.tex
\section{Background and Related Work} \label{sec:background}

This section introduces fundamental network QoS control techniques and, from this, motivates the usage of pacing.
Besides the technical basics, we describe its benefits and implementations, and present some works targeting pacing of individual TCP flows.
Finally, we summarize related works on multi-application QoE management. 
We start by defining the term enterprise network in the context of this paper.

\subsection{Enterprise Networks}

Enterprise networks are not bound to the same net neutrality laws which govern most parts of the public Internet and access to an enterprise network can be limited to approved devices.
The network operator is in full control of the applications deployed in the network and on the end-hosts. 
This is due to security concerns (e.g. malware, leakage of sensitive documents/data) and the need for performance guarantees for mission-critical applications. 
This means that end-hosts are restricted to a small set of applications, depending on the role of the employee, and that the communication of each application can be monitored.
HTTP(S) traffic passes through a proxy to perform Deep Packet Inspection (DPI) to identify sensitive documents being uploaded on an external website or malware being accidentally downloaded.

The scale can range from small businesses housed in one building to global enterprises with multiple remote campuses connected to one or multiple central offices and millions of end-hosts.
In order to adjust the available throughput according to the utility allocation, it is crucial to know or approximate the available throughput and to monitor the link utilization and packet loss.
For delay-constrained routing per application flow and load balancing, Software-defined Networking with fine-grained flow control is necessary.
If SDN control is not available, allocation can still be done based on the available forwarding graph, e.g., based on shortest-path routing.
QoS control mechanisms on the network nodes are not required, but can be used to support the overall QoS strategy.
For example, two VLANs in combination with two queues can be used to isolate managed from best effort traffic using hierarchical token bucket (HTB) scheduling.
 
\subsection{Network QoS Control Mechanisms}
On a basic level, QoS enforcement relies on two options of treating packets in the network: they either can be dropped or enqueued. 
Mechanisms that decide how packets are treated form the fundamentals of QoS control techniques, e.g., flow prioritization or rate allocation with weighted fair queuing are widely applied in today's communication networks~\cite{mirchev2015survey}. 
Table~\ref{tab:qos_mechs} summarizes and classifies the most relevant techniques and gives state of the art examples. 
In the following, we shortly describe the listed mechanisms.

Active queue management (AQM) is applied within queues of network elements and describes the intelligent drop of network packets to control the queue length~\cite{ryu2004advances}.
Excessively buffering packets causes bufferbloat and leads to increased delays. 
Random early detection (RED)~\cite{floyd1993random} is one of the well-known and widely applied mechanisms for AQM. 
Conventional tail-drop mechanisms discard all incoming packets when the queue is full.
RED drops incoming packets with a certain probability that increases with increasing queue length. 
To realize this, RED applies two thresholds: 
If the queue is (almost) empty, the probability to drop a packet is set to zero. 
If the queue is (almost) filled, all packets are definitely dropped. 
In between these two thresholds, the dropping probability increases linearly. 
That way, RED proactively prevents bufferbloat and reduces the bias of discarded packets against bursty traffic. 
Controlled Delay (CoDel)~\cite{nichols2012controlling} keeps the queuing delay of packets below a certain threshold. 
Packets are marked with the current timestamp as they enter the buffer. 
When dequeuing a packet, the CoDel algorithm computes the time it spent in the buffer. 
When the maximum delay, by default $5\,\mathrm{ms}$, is exceeded for a certain amount of time, subsequent packets are dropped at the head of the queue. 
In contrast to CoDel and RED, Explicit Congestion Notification (ECN)~\cite{floyd1994tcp} does not proactively discard packets. 
Instead, it marks packets in case of impending congestion to inform the receiver, which in turn signals the impending congestion to the sender. 
As the ECN-aware endpoints adapt the sending rate accordingly, ECN performs the queue length control and bufferbloat prevention in an indirect manner. 

Rate limiting mechanisms manage queues or flows to achieve a target traffic rate. 
One rate limiting example is policing, which controls the rate of a flow by dropping network packets. 
This is realized by applying token bucket or leaky bucket algorithms.
Tokens are created according to the target rate.
If not enough tokens are available, packets to be sent are dropped. 
In contrast to policing, where packets are dropped in case that no tokens are available, shaping enqueues packets and allows them to wait for a token to be created. 

Scheduling algorithms decide about how packets are dequeued when several queues are active. Hence, they operate between queues. 
First of all, incoming packets are classified based on pre-defined QoS policies and are accordingly inserted into one of the queues. 
The scheduling algorithm then decides about the order and frequency with which packets can be released from the different queues. 
Allowing certain queues to transmit packets more often than others, enables QoS enforcements in the sense of allocating higher bandwidth shares, i.e., different priorities to different queues. 
Such mechanisms that govern how packets are queued and de-queued, are often referred to as \textit{queueing disciplines (qdiscs)}. 
They can further be categorized as handling packets in either a classful or a classless manner. 
We omit the differentiation in Table~\ref{tab:qos_mechs}, but shortly emphasize the difference in the following.  
Classless queueing disciplines are well suited for basic traffic management and come with decreased configuration overhead compared to classful queueing disciplines. 
Classful qdiscs allow a more differentiated treatment of different kinds of traffic on the costs of increased configuration efforts, like the definition of appropriate filters and classes. 
From the examples above, Class-based Queueing (CBQ)~\cite{floyd1995link}, Hierarchical Token Bucket (HTB), and Weighted Round Robin (WRR) fall within the classful qdiscs, while Round Robin (RR) is an example for a classless qdisc.
 
%The scheduling algorithms decide about how the packets are dequeued from the active queues. 
%They classify incoming packets, e.g., based on certain QoS policies, and accordingly insert the packets into one of the queues. 
%Only one single queue can send out packets at once. 
%The scheduler decides about the frequency and number with which packets can be transmitted from the different queues.
%Attributing certain queues more often to transmit packets, allows for QoS-enforcements in the sense of allocating different bandwidth shares to queues or prioritizing certain queues. 
%Typical examples of those queueing disciplines are Class-based Queueing (CBQ)~\cite{floyd1995link}, Hierarchical Token Bucket (HTB), Round Robin (RR), or Weighted Round Robin (WRR).

The paradigm of smart queue management combines active queue management and scheduling. 
Weighted Random Early Detection (WRED)~\cite{wurtzler2002analysis} allows to apply several thresholds of dropping packets in one queue. 
For example, while packets of one QoS class are dropped if the buffer is half filled, the packets belonging to another QoS class are only dropped if the buffer is completely filled. 
Furthermore, WRED supports applying several queues with different buffer lengths. 
On the one hand, this allows to additionally influence the packet dropping probability for different QoS classes. 
On the other hand, scheduling between the queues enables realizing further QoS policies, like packet prioritization.
Flow Queue Codel (\textit{fq\_codel}, RFC8290) is an extension of CoDel.
It uses multiple queues, whereby each of the queues employs CoDel. 
A scheduler decides based on a modified Deficit Round Robin algorithm, from which queue a packet should be dequeued. 
\textit{fq\_codel} allows to enforce QoS policies by classifying the packets and allocating them accordingly to queues. 
\begin{table}
\small
\caption{Overview of traffic QoS control/allocation techniques applied in communication networks}
\vspace{-0.7em}
\label{tab:qos_mechs}
\begin{tabularx}{\textwidth}{|p{1cm}|p{1.5cm}|p{0.61cm}|p{0.61cm}|p{1.1cm}|X|p{2cm}|}
\hline
\multirow{2}{*}{\textbf{Location}} & \multirow{2}{*}{\textbf{Technology}} & \multicolumn{2}{c|}{\textbf{Action}} & \multirow{2}{*}{\textbf{Example}} & \multirow{2}{*}{\textbf{Description}} & \multirow{2}{*}{\textbf{Illustration}} \\ \cline{3-4}
& & Drop & Queue & & & \\ \hline
& \multirow{8}{1.5cm}{\textbf{Active queue management:} Manage the queue length} & \multirow{6}{0.55cm}{X}   &  & \multirow{3}{0.9cm}{RED} & Drops packets based on statistical probabilities instead of conventional tail drop. Prevents high delays resulting from full buffers. & \raisebox{-\totalheight}{\includegraphics[width=0.12\textwidth]{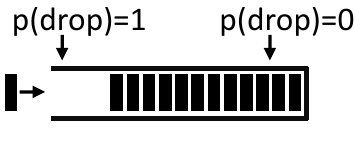}} \\ \cline{5-7}
 & &   & & \multirow{3}{0.9cm}{CoDel} & Reduction of packet transmission delays by preventing large and constantly full buffers & \raisebox{-\totalheight}{\includegraphics[width=0.15\textwidth]{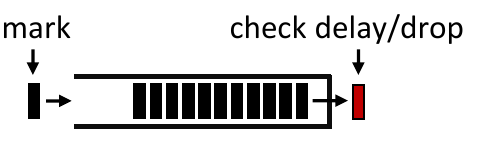}}\\ \cline{3-7}
 \multirow{6}{1cm}{Within queues} & &   & & \multirow{3}{0.9cm}{ECN} & Notification about network congestion without dropping packets & \raisebox{-\totalheight}{\includegraphics[width=0.1\textwidth]{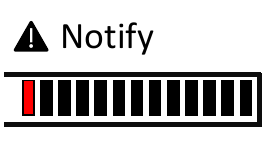}} \\ \cline{2-7}
& \multirow{8}{1.5cm}{\textbf{Rate limiting:} Achieve target traffic rate} &\multirow{4}{0.55cm}{X} &  & \multirow{4}{0.9cm}{Policing}   & Tokens are created with a rate corresponding the target traffic rate. If no tokens are available, incoming packets are dropped. & \raisebox{-\totalheight}{\includegraphics[width=0.149\textwidth]{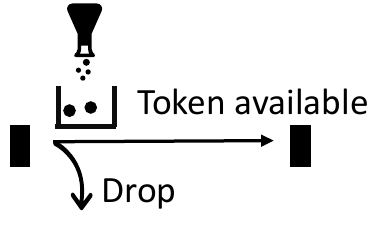}}\\ \cline{3-7}
& & & \multirow{3}{0.55cm}{X}  & \multirow{4}{0.9cm}{Shaping} & Tokens are created with a rate corresponding the target traffic rate. If no tokens are available, incoming packets are enqueued. & \raisebox{-\totalheight}{\includegraphics[width=0.12\textwidth]{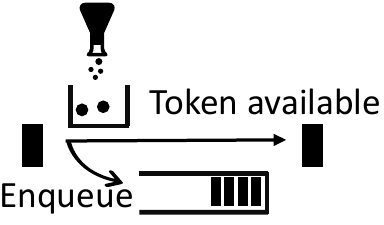}}\\ \hline
&&    & \multirow{3}{0.9cm}{X} & \multirow{3}{0.9cm}{RR} & Round Robin lets every active data flow take turn in transferring packets on a shared channel in a periodically repeated order. & \\ \cline{3-6}
\multirow{4}{1cm}{Between queues}&\multirow{5}{1.5cm}{\textbf{Scheduling:} Allocate resource to queues}&  & \multirow{3}{0.61cm}{X} & \multirow{3}{0.9cm}{CBQ}  & Divides user traffic into a hierarchy of classes and performs class based queueing so to allocate bandwidth to traffic classes. & \multirow{4}{2cm}{\raisebox{-\totalheight}{\includegraphics[width=0.15\textwidth]{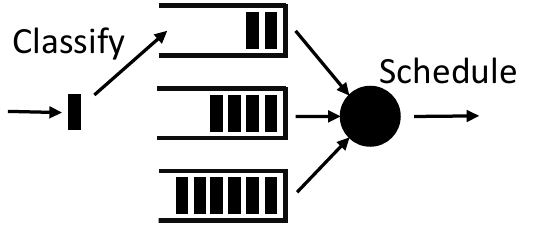}}} \\ \cline{3-6}	
&&    & \multirow{3}{0.9cm}{X} & \multirow{2}{0.9cm}{WRR}  & Allows to differentiate QoS classes by allowing certain queues to put more packets on the wire.   & \\ \cline{3-6}
&&    & \multirow{2}{0.9cm}{X} & \multirow{2}{0.9cm}{HTB}  & Hierarchical token bucket allows for setting bandwidth thresholds to different flow classes. & \\ \hline

\multirow{6}{1cm}{Hybrid: Within and between queues} & \multirow{6}{1.5cm}{\textbf{Smart queue management:} QoS-aware queue mangagement} & \multirow{3}{0.9cm}{X} & \multirow{3}{0.9cm}{X} & \multirow{3}{0.9cm}{WRED} & Supports several queues that vary in buffer size and allows several thresholds per queue. 
Packets of higher prioritized flows are less likely to be dropped. & \raisebox{-\totalheight}{\includegraphics[width=0.15\textwidth]{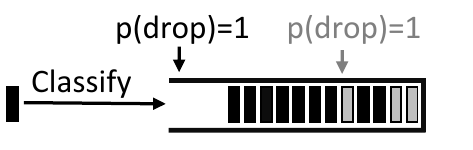}} \\ \cline{3-7}
& & \multirow{3}{*}{X} & \multirow{3}{0.9cm}{X} & \multirow{3}{0.9cm}{FQ-CoDel} & Flow queue CoDel (fq\_codel) extends CoDel by applying several queues. Allows for differentiating QoS classes. & \raisebox{-\totalheight}{\includegraphics[width=0.15\textwidth]{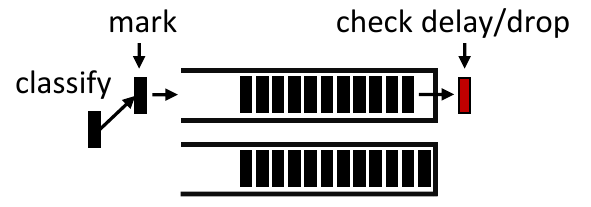}}\\ \hline
\end{tabularx}
\Description[Comparison of rate limiting, scheduling and queue management.]{Quality of service techniques for within queues (active queue management, rate limiting), between queues (scheduling) and combination of both (smart queue management).}
\end{table}

Although many QoS enforcement mechanisms exist and are applied in today's networks, they cannot be straightforwardly applied in our case. 
Some of the techniques listed in the table are not powerful enough. 
ECN, for example, is capable to influence the sender's rate to prevent packet loss, but does not allow to set a specific rate. 
The major drawback when it comes to applying those mechanisms is the limited number of configurable queues in network elements such as switches and routers.\footnote{Jim Warner, \url{https://people.ucsc.edu/~warner/buffer.html}, last accessed: 11.10.2018}
As a consequence, QoS can only be enforced on aggregated flows and QoS classes. 
Hence, the limited scalability hinders a fine-granular QoS control.  
Shifting the QoS enforcement from network nodes to the end hosts constitutes a scalable method that allows for fine-grained QoS control. 
For that reason, we propose to apply TCP pacing to enforce traffic rates on a per application basis. 

\subsection{Pacing}

The term pacing is used in different contexts and it is important to distinguish where it is applied and who is dictating the pacing rate.
There can be pacing per interface, per application and per network socket or a combination of all three.
The pacing rate can be set autonomously, e.g., like in TCP pacing, or by an external entity, e.g., by a central network controller.
The term TCP pacing is an example where the rate is set autonomously and refers to the technique where the packets of one TCP transmission window are spread out over the measured RTT \cite{aggarwal2000understanding}.
The target pacing rate is determined by the congestion control algorithm based on observed packet loss or delay.
In this work, when we use the term pacing, we apply pacing per-application flow and it is set by the central network controller.
One application flow can include one or multiple streams (TCP) or datagram-based (UDP) transmissions which share the same source, destination and network path.
Hence, all packets sent by the sockets of an application flow have to share the allocated pacing rate.
In the following, we shortly introduce the pacing implementation of the Linux Kernel.
Afterwards, we highlight the advantages of this technique compared to other rate limiting approaches, i.e., policing and shaping. 
Finally, other works relying on pacing are summarized.

\subsubsection{Pacing Implementations}

Pacing follows the approach of placing gaps between outgoing packets so to evenly space data transmissions~\cite{aggarwal2000understanding,cheng2016making}. 
In the case of the Linux pacing implementation, the departure time of the next packet $time\_next\_packet$ is determined by the current time $now$, the size of the current packet $pkt\_len$, and the target pacing rate $target\_rate$: 
\[
time\_next\_packet = now + \frac{pkt\_len}{target\_rate}
\]

For details on the technical fundamentals and the way pacing is applied in this work, please refer to Section~\ref{subsec:pacingimpl}.
Google is currently putting much efforts in developing efficient, rate-compliant, and scalable traffic control mechanisms, mainly for a deployment in data centers. 
To do so, they implement pacing in many of their recent approaches.  
With \textit{TIMELY}, they propose an RTT-based congestion control~\cite{mittal2015timely}. 
Their congestion-based congestion control (BBR)~\cite{cardwell2016bbr} is implemented in the Linux kernel and used by all Google and YouTube server connections.
With \textit{Carousel}~\cite{saeed2017carousel} they present a scalable traffic shaping mechanism by controlling packet release times where the target rate can be set by external entities per traffic class.
As we do not have those strict requirements on scalability as for \textit{Carousel}, we apply a custom version of the Linux \textit{fq} implementation (Section~\ref{subsec:pacingimpl}).

\subsubsection{Benefits and drawbacks of pacing}
Pacing eliminates several drawbacks of other strategies for traffic rate control. 
While policing drops packets exceeding the target rate and shaping enqueues those packets, pacing follows the approach of delaying packets so to reach a certain rate. 
On the one hand, this eliminates the problem of increased overall network load resulting from retransmitting dropped packets of policed flows. 
On the other hand, there is no RTT inflation, as with shaping. 
Policing interacts poorly with TCP, as a result, policed flows suffer from low throughput even at low packet loss rates~\cite{flach2016internet}. 
In contrast, pacing can increase the link utilization in shared environments. 
Delaying the outgoing packets at the sender in a controlled manner reduces burstiness, which implicates less packet loss and results in fewer triggers of TCP's congestion control.
Furthermore, configuring target rates at end hosts brings the advantage of scalability, compared to other techniques.  
%Due to the limited number of configurable queues in network nodes, shaping and policing only allow for a coarse-grained QoS control on flow aggregates. 
By shifting the QoS control to the involved end-hosts, pacing facilitates a fine-grained control on flow- and application-level.

However, studies have also shown that this only applies to some cases, while even with all-paced TCP flows the performance is worsened in many cases \cite{aggarwal2000understanding,ghobadi2013tcp}. 
According to the authors of the studies, this can be attributed to the fact that TCP pacing delays the congestion signal and that pacing results in synchronized packet drops.
A pacing system that shapes traffic under consideration of the buffer queue is proposed in~\cite{cai2010practical}.
The authors introduce Queue Length Based Pacing (QLBP) to shape the traffic at access networks, so to smooth the traffic before entering the core network. 
This is especially interesting for small buffer networks, where packet loss is more likely to occur.
The reduced packet loss, as a consequence of the decreased burstiness, results in a nearly fully link utilization when using the proposed solution. 
The QLBP algorithm is also applied in~\cite{cai2011study} to study the impact of pacing on different network traffic conditions. 
The authors conclude that pacing is especially beneficial in networks with small buffers, where packet loss, as a result from bursty traffic, can significantly reduce network performance. 
They furthermore show that pacing can have a small negative impact on short-lived flows, if the parameters are not set appropriately. 
Finally, it is shown that the fairness achieved by pacing only slightly differs from the fairness as achieved by TCP. 
The performance of host traffic pacing and edge traffic pacing, i.e., pacing the traffic before it enters the core network, is compared for small buffer networks in ~\cite{gharakheili2015comparing}. 
The results indicate for most of the evaluated scenarios that edge pacing performs at least as well as host pacing in terms of link utilization. 
Edge pacing also has practical benefits, as it does not require an adaptation of the involved clients.
%Google researcheres introduced Carousel~\cite{saeed2017carousel}.
%It addresses the scalability of end-host pacing in data centers. 
%The proposed framework improves CPU-utilization while simultaneously obtaining a better rate conformance. 
%This is achieved by applying the following three techniques.
%Firstly, Carousel relies on a single queue system instead of using one queue per rate limit. 
%This is implemented by timestamping each packet, before putting it into the queue which is realized as a timing wheel. 
%The packets are released according to their send-time timestamps. 
%Secondly, it implements deferred completions. Instead of sending the completion signal, which allows the kernel stack to send more packets, when a packet leaves the networking stack, 
%the signal is deferred until Carousel dequeues the packet to the wire. 
%With this backpressure mechanism, head of line blocking and memory pressure are reduced. 
%Thirdly, the allocation of solely one single queue shaper to one CPU core enhances efficiency, as overhead for locking and synchronization can be economized. 
A critical analysis on pacing is performed in~\cite{wei2006tcp}. 
The authors evaluate the impacts of pacing for several TCP implementation and scenarios. 
They conclude that the benefits when applying pacing depend on the used TCP implementation and on the performance metrics that are relevant for a specific application. 
However, due to the tendency to high speed protocols, they predict an increasing motivation to use pacing in future. 
Furthermore, they showed that in some cases, pacing was capable to improve the performance of both, paced and un-paced flows. 
As a drawback, the work highlights the unfairness among paced and non-paced flows in terms of bandwidth, as paced flows do not receive their fair share when competing with non-paced flows. 

\footnotetext[4]{\url{https://lwn.net/Articles/564825/}}

\begin{table}[t]
\caption{Summary of state-of-the art approaches that make use of TCP pacing, along with their scope and the entity deciding about the pacing rate.}
\label{tab:pacing}
\begin{tabularx}{\textwidth}{|p{1.63cm}|p{4cm}|p{4cm}|p{4cm}|}
\hline
%\begin{tabular}{llll}
\textbf{Technique}               & \textbf{Category}                  & \textbf{Scope}                                         & \textbf{Rate set by} \\ \hline
BBR~\cite{cardwell2016bbr}, TIMELY~\cite{mittal2015timely}  & TCP Pacing                         & One TCP socket                                         & TCP congestion control \\ \hline
Carousel~\cite{saeed2017carousel}  & Efficient and scalable pacing implementation    & Flexible, based on traffic classes, evaluated per flow & External controller or TCP congestion control \\ \hline
FQ\footnotemark[4] & Linux kernel pacing implementation & Per flow                                          & Primarily congestion control algorithm, can be manually overwritten \\ \hline
BwE~\cite{kumar2015bwe}                              & Hierarchical bandwidth allocation & From global to per computing task                      & External controller \\ \hline
QLBP~\cite{cai2010practical,cai2011study} & Edge pacing                        & Single queue per interface                             & Adaptive based on queue-length \\ \hline                                     
Our work                         & End-user application pacing                 & Per a subset of application's sockets                  & External controller   \\ \hline                                             
\end{tabularx}
\Description[BBR, Carousel, FQ focus on socket and flow, BwE on global bandwidth allocation, this work on applications.]{}
\end{table}

We summarize the approaches that rely on TCP pacing in Table~\ref{tab:pacing}.
%Current research either focuses on pacing from a network-centric point of view or observes technical aspects like CPU utilization. 
It shows that pacing is applied on per-interface, per-flow, and per-task levels, but so far not on a per-application level.
Although these approaches might provide fairness on a per-flow level, they do not provides fairness between applications (10 connections opened by a web browser vs. 1 connection for a file download) and it still remains unclear how this could be applied to UDP-based applications where there is no operating system support for TCP-style probing of the available throughput.
This work aims at closing the gap of considering pacing from an application-centric perspective, i.e., to evaluate its feasibility for application-aware network management.
We investigate the conformance of actual rates and delays to the target values, which dictates the degree of granularity to which QoE can be controlled. 
As we will find that pacing constitutes a feasible method to do so, we present a proof-of-concept architecture for optimizing QoE fairness in a multi-application environment.

\subsection{Related Work on Multi-Application QoE Management}
\begin{table}
\small
	\caption{Overview of related works targeting multi-application QoE-awareness and their classification in terms of utility function, determination of QoE-aware resource shares, and allocating of determined resources. 0~denotes that no utility functions are applied at all, +~denotes that utility functions are applied mapping either AQoS or AQoS plus NQoS to QoE, ++~represents utility functions solely relying on configurable network resources, e.g. bandwidth or PSNR in radio access networks.}
	\vspace{-0.7em}
	\label{tab:multiapplication_appraoches}
	\begin{tabularx}{\textwidth}{|p{0.7cm} | p{0.8cm} | p{3.2cm} | p{4.0cm} | X |}
	\hline
	\multirow{3}{*}{\textbf{Source}} & \multicolumn{2}{c|}{\textbf{Utility function}}  & \multirow{3}{*}{\textbf{Determine}} & \multirow{3}{*}{\textbf{Allocate}} \\ \cline{2-3}
	 & Classi- & \multirow{2}{*}{Description} & & \\ 
	& fication& & &   \\ \hline
	
	\cite{gomez2013towards} & + & Mapping AQoS to QoE & Not specified  & Generic control concepts \\ \hline
	\cite{tang2014qoe} & ++ & Mapping NQoS to QoE & Particle swarm optimization (PSO) based algorithm  & Applying the proposed algorithm to resource block allocation technique in LTE \\ \hline
	\cite{sacchi2011qoe} & ++ & Mapping NQoS to QoE & Game theoretic approach & Radio resource management applying proposed game theoretic approach   \\ \hline	
	\cite{liu2012novel} & + & Mapping NQoS and AQoS to QoE & Optimization based on multi-choice knapsack problem (MCKP) & Carrier scheduling applying proposed optimization algorithm  \\ \hline
	\cite{fei2015qoe} & ++ & Mapping of network bandwidth to QoE & Solving multi-objective optimization problem & Joint subcarrier and power allocation scheme \\ \hline
  \cite{ferguson2013participatory,ferguson2012participatory} & 0 & Application feedback instead of utility functions & Not specified & Admission Control, bandwidth guarantees\\ \hline
	\cite{salles2005fair} & 0 &  Hypothetical utility functions mapping NQoS to QoE & Proposed algorithm optimizing bandwidth allocation & WFQ scheduling with QoE-optimized weights  \\ \hline
	\cite{georgopoulos2013towards} & ++ & Mapping screen resolution and bitrate to SSIM & Branch and bound algorithm to find optimal set of video bitrates& Video bitrate guidance for heterogeneous clients 
  \\ \hline
	\cite{kumar2015bwe} & + & Mapping bandwidth to an arbitrary fair share & Novel Multi-Path Fair Allocation (MFAA) algorithm & Enforced via pacing at the hosts
  \\ \hline
	\end{tabularx}
\end{table}

Several efforts have been made towards QoE-awareness in multi-application scenarios. 
Some relevant approaches are summarized in Table~\ref{tab:multiapplication_appraoches}. 
The first column denotes the investigated approaches. 
The remaining table columns represent the three challenges introduced beforehand: \textit{define}, \textit{determine}, and \textit{allocate}. 
However, we found that none of the reviewed work explicitly defines the required resources to obtain a certain 
Mean Opinion Score (MOS)
\footnote{ITU-T Recommendation P.800. Methods for objective and subjective assessment of quality},
but they all utilize in some form utility functions that map application QoS (AQoS) and/or network QoS (NQoS) to express QoE. 
The MOS scale describes the experience of a user with the application on a scale of one to five where the scale is labeled with \{Bad, Poor, Fair, Good, Excellent\}.
For that reason, we replaced in the table the \textit{define}-step by a classification and a short description of the applied utility function.
In the following, when reporting on related work, we focus on how and which utility functions have been applied, how the appropriate resource shares are determined, and the applied methods to allocate the resources. 

BwE\cite{kumar2015bwe} introduces a global hierarchical top-down bandwidth allocation scheme used in Google's internal network for distributed computing tasks. 
Bandwidth allocation is done via a function that maps bandwidth to a "relative priority on an arbitrary, dimensionless measure of available fair share capacity".
The BwE reference is important as it shows that global and large-scale bandwidth allocation is indeed possible in production environments. 
But how to derive an allocation for end-user applications and how they benefit from it, is not discussed in BwE.
In contrast, this paper at hand focuses on end-user applications and the interplay with, and possibilities for, network control to guarantee a specific user experience to the end users. 

Many related works with a focus on multi-application QoE-aware networking are associated to the mobile domain~\cite{gomez2013towards,tang2014qoe,sacchi2011qoe,liu2012novel,fei2015qoe}. 
Several KPIs are proposed to be monitored at network elements in the architecture of~\cite{gomez2013towards}, including packet loss rate, throughput, and RTT. 
At the clients, network-related parameters, e.g., delay, and application-based metrics including web page download time or video buffer and bit-rate can be measured. 
The collected AQoS metrics are used to estimate the per-application QoE using models from literature. 
One of the presented use-cases in~\cite{gomez2013towards} considers a QoE optimization based on the estimated QoE values. 
To do so, the authors list a variety of parameters that can be configured along the protocol stack in order to control QoE. 
The work does not provide a specific algorithm for determining the required resources, nor does it propose a designated method for allocating them. 
Instead, the authors outline several possible control actions like bandwidth limiting or QoE-aware capacity planning. 
As the applied utility functions only rely on AQoS, the QoE can only be controlled in a qualitative manner, meaning enhancing and degrading the QoE, but not controlling it so to achieve a specific MOS value.

Tang, et al.~\cite{tang2014qoe} proposes a novel algorithm for resource block allocation in LTE systems to maximize QoE whilst preserving fairness among users. 
The authors also use existing models to estimate user QoE, but adapt the models so to express the mean opinion score (MOS) solely from network-parameters like delay or packet error probability. 
Based on these models, the authors present a resource block allocation algorithm that is based on Particle Swarm Optimization. 

Another QoE-aware resource scheduling algorithm for mobile networks is based on a game-theoretic approach~\cite{sacchi2011qoe}.
The QoE is estimated for various applications using models from literature that map network parameters to MOS.
The users' data flows cooperate with each other in a proactive manner and jointly optimize the QoE in a game-theoretic based manner. 
Instead of using the conventional throughput maximizing algorithm in radio resource management of OFDMA, the authors propose to implement their scheduling algorithm which aims on maximizing the fairness among heterogeneous users. 

The approach described in~\cite{liu2012novel} targets QoE-awareness in mobile LTE-Advanced networks. 
In the QoE modeling step, both NQoS and AQoS are used for estimating the user perceived quality for different application types.
The estimated QoE and available bandwidth are inputs to the resource scheduling algorithm, which solves a multi-choice knapsack problem (MCKP) that maximizes the sum of all users' MOS values. 
The component carriers are dynamically scheduled according to the network traffic load by this QoE-aware scheme.

A further approach towards QoE-driven resource allocation in wireless networks is~\cite{fei2015qoe}. 
The authors apply utility functions which express MOS for various applications as functions of different NQoS parameters. 
Thereby, they assume a packet error probability of 0, a packet loss rate of 0, and fixed frame rate in the case of video streaming applications.
Using these simplified utility functions, the authors propose a solution to a multi-objective optimization problem which aims at maximizing MOS. 
As network resource control mechanisms, the authors apply an efficient allocation of subcarriers among the active users. 

The concept of Participatory Networking is proposed in~\cite{ferguson2013participatory,ferguson2012participatory}.
It describes an API that can be used by applications, end-hosts, and devices to interact with the network. 
A centralized controller is authorized to delegate read and write access to the network participants. 
Using the write access, applications, users or end-hosts can reconfigure the network according to their needs and can provide knowledge to the network, e.g., their future traffic demands. 
Hence, no utility functions that map AQoS or NQoS to QoE are needed, as the application instances directly communicate their requirements to this controller. 

\cite{salles2005fair} applies hypothetical piecewise linear functions that map bandwidth to QoE and propose a new scheduler for fair and efficient bandwidth allocation in shared networks. 
Using these utility functions, they optimize the bandwidth per flow so to have a fair utility over all active applications. 
According to the bandwidth shares, the weighted fair scheduler allocates respective weights to the flows. 
Simulation results show that the minimum utility can be increased significantly, while maintaining the same average utility in most of the cases, compared to a conventional max-min-fairness approach. 

\cite{georgopoulos2013towards} presents an SDN-based framework to support a fair video QoE for all clients within a shared network. 
The utility function maps a client's device resolution and bitrate to structural similarity (SSIM)~\cite{wang2004image}. 
Considering the current network capacity, a controller decides about the bitrate for each video client, so to provide a similar quality to each of them. 
The bitrates are communicated to the streaming clients, which in turn request the respective quality level from the video content server. 

The presented strategies are all steps towards QoE-awareness in multi-application systems. 
Some of the works rely on state of the art control mechanisms, but propose novel resource scheduling or allocation techniques. 
However, the applied utility functions often depend on features, which cannot be influenced in a direct manner. 
As a result, those approaches allow for a qualitative, less targeted QoE control. 
For example, a low video quality implies a low MOS value. 
Providing more bandwidth will enhance the playback quality and increase MOS, but it is not possible to quantify the impact of providing a certain amount of bandwidth on MOS scale. 

We present an approach that allows to quantitatively map the NQoS parameters bandwidth and delay to MOS. 
Furthermore, we propose to apply network pacing, which allows us to control both, the bandwidth allocated to a flow and the end-to-end delay. 
Having utility functions which only rely on controllable parameters allows for a targeted, fine-grained QoE optimization.

%% file: content/04_overview_architecture.tex
\section{System Design} \label{sec:overview_architecture}  
The background on multi-application QoE architecture designs shows that previous proposals cannot combine accurate identification of application- and user-aware resource demands with scalable resource allocation.
In the following we propose a new design considering the following aspects:
\begin{enumerate*}[label={\alph*)}]
\item Awareness of the active applications in the network and their demands,
\item per-application resource allocation and 
\item per-application forwarding for delay- and capacity-constrained routing.
\end{enumerate*}
Our design relies on a centralized control, whose major drawback is the introduction of a single point of failure (SPOF). 
Although not addressed in detail in our work, we would like to emphasize that several approaches exist for physically distributed, but logically centralized SDN control planes, to overcome the SPOF problem.
Proposed solutions either rely on hierarchical~\cite{hassas2012kandoo} or flat organizations~\cite{tootoonchian2010hyperflow} and enable scaling the load among several controllers.

%\subsection{Overview}

\begin{figure}[t]
\centering
\includegraphics[width=340pt]{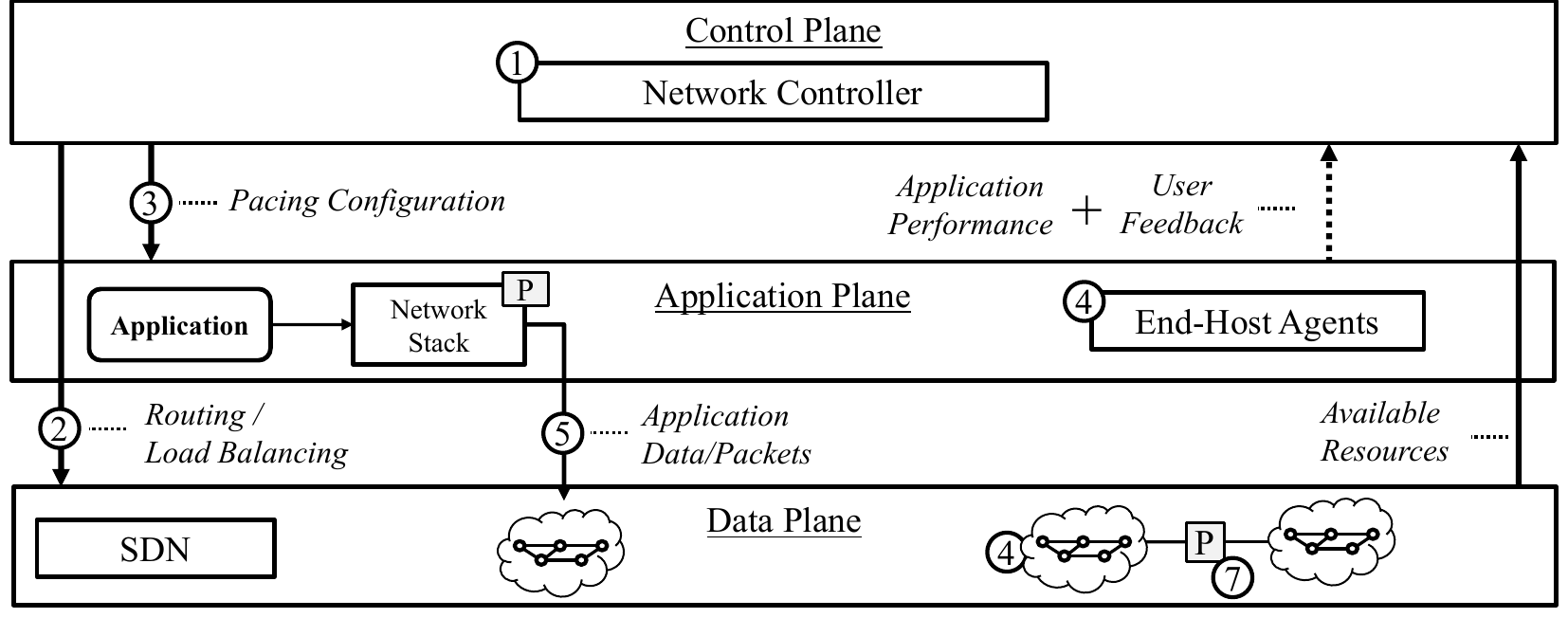}
\vspace{-0.7em}
\caption{
Overall system design. 
A logically centralized network controller \protect\fkey{1} provides per-application delay-constrained routing and resource allocation.
Applications are identified by software agents at the end-hosts \protect\fkey{4} and resources are allocated through restricting the total sending rate of applications at the hosts' networking stacks (\protect\fkeynb{3}, \protect\pacer{}) and at the network edges \protect\fkey{7}.
Per-application routing is implemented by using SDN protocols to push individual forwarding rules to the network devices \protect\fkey{2}.
Per-application delay requirements are fulfilled by a careful selection of the flow path and target link utilizations.
}
\label{fig:system_overview}
\Description[Control plane on top of the application plane, on top of data plane.]{Control plane with network controller, application plane with end host agent, application and network stack. Data plane with software-defined devices and pacers on network links.}
\end{figure}

Figure \ref{fig:system_overview} illustrates the system design.
A logically centralized network controller \fkey{1} exerts control over the forwarding devices, the \textit{data-plane}, by applying network control \fkey{2} through SDN protocols.
SDN protocols, such as OpenFlow, enable per-flow routing by pushing simple match-action rules to SDN-enabled devices.
Resources are allocated in the network by pacing (\rectangled{P}) the data-rate sent by traffic sources into the network \fkey{5}.
Pacing is applied at the edges of the network, i.e., at end-hosts such as clients and servers, or at gateways \fkey{7}. 
Software agents on the end-hosts \fkey{4} allow the network controller on the one side, to know about which applications access the network, and on the other side, to apply pacing to the applications \fkey{3} at the host's networking stack \pacer{}.
Applying pacing at the end-hosts' networking stacks can support tens of thousands of individual flows with low additional resource consumption for the host \cite{saeed2017carousel}.
All conversations in the network are subject to the pacing set by the network controller.
If a conversation cannot be paced at its networking stack, pacing can then be applied, for example, at the first hop in the network.
Delay requirements are fulfilled by selection of appropriate links and target link utilizations.
We describe how pacing is implemented in our set-up as part of the experiment design in Section \ref{subsec:pacingimpl}.

\subsection{Intents and Intent Hierarchy}

%\noindent
%\textbf{Intents}
Some application flows transported on the network, such as system monitoring or building surveillance, have predictable traffic patterns and determining a suitable data-rate to allocate is straightforward.
Furthermore, periodic background jobs, like backup data transfers, can be scheduled based on the approximate amount of data and the deadline for completion.
However, for user-facing applications, the variety of demands is higher.
Determining the appropriate pacing data-rate for such applications is challenging.
It is insufficient to consider only the \textit{class} of an application, e.g., web browser, but also for what purpose the application is used.
For example, modern web browsers are an execution environment for a variety of business applications, from employee and financial management to video streaming (\textit{DASH}, \textit{HTMLMediaElement}) and video conferencing (\textit{WebRTC}).
Hence, we distinguish between application \textit{classes} and application \textit{intents}.
An intent can be specific, such as a video stream of a surveillance camera with specific encoder settings, or broad, such as general web browsing.
A running application can also participate in multiple conversations with different intents and conversation endpoints.
Hence the resource demands of a conversation are defined by the tuple of \textit{(class, intent)}.
%With \textit{(application, intent)} with denote the demand of a specific application instance.

\begin{figure}[t]
\centering
\includegraphics[width=425pt]{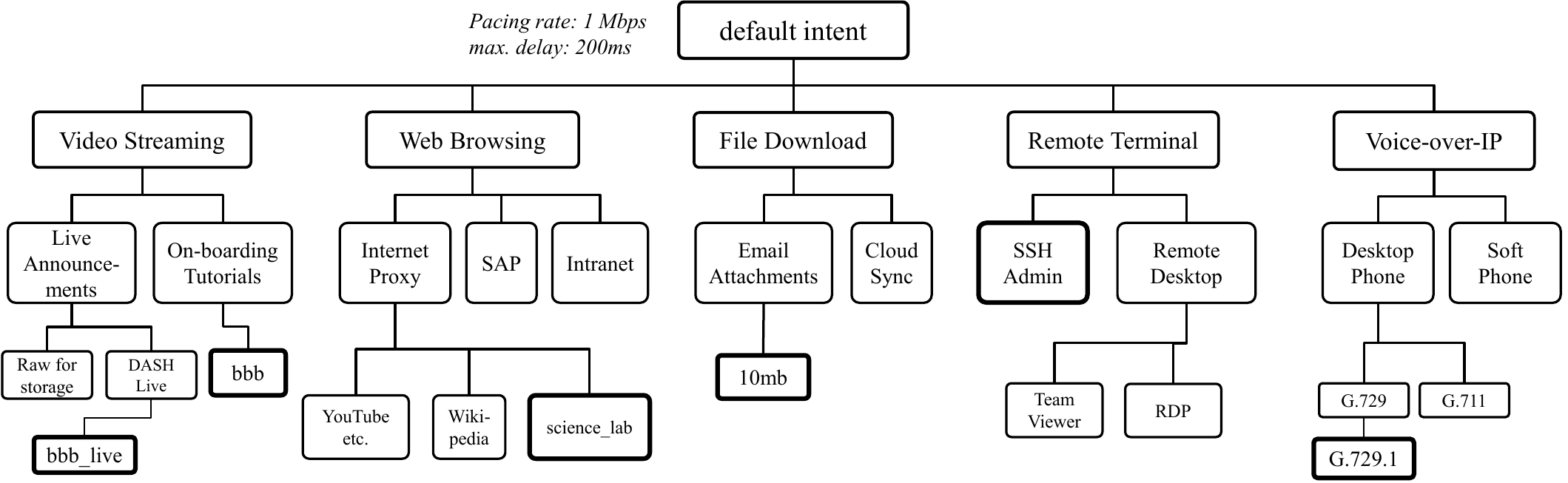}
\vspace{-0.6em}
\caption{
Illustration of a possible hierarchy of application classes and intents. 
Classes and intents specify the utility function and user experience model to use for the target utility calculation and resource allocation.
There is a trade-off between predictable application QoE and the effort for the company to construct the hierarchy.
If an application's intent cannot be identified, fall-back rules can be applied to select a higher-level intent with the cost of reduced QoE accuracy.
Highlighted intents are part of the evaluation.
}
\label{fig:intents}
\Description[default intent as root of tree with application classes underneath.]{Default intent provides 1 Megabit per second as root of a tree with video streaming, web browsing, file download, remote terminal and VoIP underneath.}
\end{figure}

%\noindent
%\textbf{Intent Hierarchy}
Identifying the intent of an application accurately enables the specification of precise application demands and the selection of suitable user experience models.
Both are essential to implement predictable application performance and to improve accuracy in terms of QoE for the user.
We argue that in an enterprise deployment, a holistic identification of all classes and intents is infeasible. 
Therefore, we propose a hierarchy of intents as illustrated by Figure \ref{fig:intents}.
The figure shows a possible enterprise intent hierarchy by example.
At the root of the hierarchy, there is a default intent which offers basic guarantees in terms of throughput and delay to unidentified applications.
The root intent is followed by the application classes such as video streaming and remote terminal work.

Intents can be specified with an arbitrary hierarchy depth.
If an application's intent cannot be identified accurately, a higher-level intent can be selected.
However, this comes with the cost that the allocated resources do not fit to the targeted application performance.
For example, the hierarchy in the figure specifies the two common voice codecs G.729 and G.711 as sub-intent for desktop VoIP-phones.
If the codec is known, e.g., based on the MAC/IP address of the phone or from a database, the demand and user experience model are well-defined.
If the conversation from the phone can only be identified as a desktop phone, one may define the highest known demand from all codecs.
How to create such a deep hierarchy is out of scope of this paper. 
We restrict our hierarchy to the five classes and six intents as highlighted in bold in the figure.
One can imagine that a combination of user-feedback and network/application monitoring, combined via machine-learning and some manual work, results in an accurate representation of the enterprise environment.

\subsection{Network Controller and Application-Awareness}

The question remains how applications can convey their class and intents to the network controller.
We propose local agents \fkey{5} at the end hosts as an interface between applications and network control.
There are two basic options from here.
The first one is to modify the applications to report their identity and intent(s) to the agent. 
This can be done with a standardized API, for example through client or server extensions.
The agent then forwards this information to the network controller and waits for the controller to decide on the appropriate pacing rate to apply.
The second option could be for the agent to monitor connection establishment or perform Deep Packet Inspect (DPI) and classify the conversations by matching it to known endpoints, header fields, packet payloads, process names or function calls.
Other techniques from the area of application performance management (APM), like code injection or tracing in the operating system, could also be used.

Without installing an agent, you could also capture packets of a conversation at the first hop, for example by using the SDN protocol OpenFlow and its \textit{packet\_in} feature \footnote{http://flowgrammable.org/sdn/openflow/message-layer/packetin/, last accessed: 11.10.2018}, or by custom middle-boxes.
However, in the network most traffic is encrypted and identification becomes difficult.
There is no one-size-fits-all solution for how to identify applications and their intents as the available options depend on the specific enterprise environment.
One can expect that a combination of rule- and pattern-based matching and machine-learning reduces the required manual work to a minimum.

%For our proof-of-concept environment we deploy a custom agent and modify the deployed applications.
%\fixme{[AB: Trotzdem generalisierbar?]}

\subsection{Utility Functions}

Once the application class and intent of a conversation are identified, the controller looks up the utility function for the \textit{(class, intent)} tuple from a database.
The utility function describes the relationship between demand, in terms of minimum throughput and maximum delay, and benefit, in terms of utility.
%We define demand as a minimum throughpt requirement and maximum tolerated delay for a conversation.
We define utility as a dimensionless unit in the range of $[1, 5]$, which describes the satisfaction of the user with the service. 
The subsequent Section \ref{sec:applications} introduces the utility functions in detail.

%\subsection{Switch Delay \& Queuing Model}
%\begin{figure}[h]
%\centering
%\includegraphics[width=0.97\columnwidth]{figures/04_delayapprox.pdf}
%\caption{Delay (M/M/1 system)}
%\label{fig:delayapprox}
%\end{figure}

%\begin{itemize}
%\item \fixme{sollte hier nicht auch der Fairness Begriff definiert werden (kommt später nur mehr als MO Problem Formulierung, das ist nicht intuitiv verständlich), insbesondere, wie er von den utilities abhängt etc.}
%\item \fixme{Kap. 3 beinhaltet sonst eben nur die intents und utilities, da fehlt etwas}
%\item \fixme{enterprise network?}
%\end{itemize}

%% file: content/05_application_utility.tex
\section{Utility Function Definition} \label{sec:applications}

Comparing the performance of different applications with conceptual different KPIs requires mapping functions to a common scale.
We denote the scale as \textit{user-aware utility scale} and we define it with a dimensionless quantity in the range of [1, 5].
The utility functions then describe the relationship between the amount of resources allocated to an application and the resulting experience of the user with the application.
In the following section we define the utility functions for selected classes of applications and intents.
First, we present the considered application classes, intents, and KPIs of the deployed implementations.
Second, we discuss the selected user experience models from the literature.
Third, we define the utility functions based on measurements and the user experience models.
%\fixme{[AB: was ist neu hier?]} <- hopefully answered by related work

We consider five application classes: 
Web browsing, file download, video streaming, remote terminal work, and Voice-over-IP (VoIP) (Table \ref{tab:applications}).
\textit{Web browsing} covers a wide range of use cases, as modern web standards facilitate the move from proprietary and platform-dependent software to responsive web applications running in the browser.
\textit{File download} is the batch-transfer of data the user is waiting for, such as an email attachment.
Use cases for adaptive \textit{video streaming} in the enterprise range from announcements to training videos, such as on-boarding lectures for new employees.
Depending on the purpose, both, video-on-demand as well as live transmissions, are conceivable.
In particular major announcements are taxing for the infrastructure when viewed by a large fraction of the staff in a short time-frame.
\textit{Remote terminal work} by secure shell access allows administrators to access the terminals of servers, hosts, and switches from anywhere.
The application class \textit{VoIP} includes office phones, conferencing by software or in the browser, and VoIP applications on smartphones.
We denote the combination between an application class and intent as application \textit{type} and use the types \web{}, \dl{}, \voip{}, \live{}, \ssh{} and \voip{} as shorthands for the investigated combinations of application classes and intents.

\begin{table}[t]
\centering
\caption{Applications, Intents and Key Performance Indicators}
\vspace{-0.5em}
\label{tab:applications}
\begin{tabular}{llllll}
\hline
\multicolumn{1}{l}{\textbf{Class}} & \multicolumn{1}{l}{\textbf{Application}} & \multicolumn{1}{l}{\textbf{Intent(s)}} & \multicolumn{1}{l}{\textbf{Shorthand(s)}} & \multicolumn{1}{l}{\textbf{KPI(s)}} & \multicolumn{1}{l}{\textbf{QoE Model}} \\ \hline
Web Browsing                       & Firefox, selenium\textsuperscript{\ref{fn:selenium}}                 & \textit{science\_lab} & \web{} & Page Load Time & Egger et al. \cite{egger2012time} \\ 
File Download                      & Python \textit{requests}          & \textit{emailattach} & \dl{} & Download Time & Egger et al. \cite{egger2012time} \\
%Video Streaming                    & TAPAS \cite{tapas}                       & \textit{bbb}, \textit{bbb\_live} & \vod{}, \live{} & Average Quality, (+ Stallings, Switches) \\
Video Streaming                    & TAPAS \cite{de2014tapas}                       & \textit{bbb}, \textit{bbb\_live} & \vod{}, \live{} & Average Quality & \textit{custom} \\
Remote Terminal                    & SSHv2, paramiko\textsuperscript{\ref{fn:paramiko}}  & \textit{sshadmin} & \ssh{} & Response Time & Casas et al. \cite{casas2013quality} \\
Voice-over-IP                      & D-ITG \cite{BottaDP12}         & \textit{g729.1} & \voip{} & Delay, Loss, (+ Jitter) & Sun et al. \cite{sun2006voice} 
\end{tabular}
\vspace{-0.5em}
\end{table}

\subsection{Applications, Intents and KPIs}

Next we discuss the implementations, KPIs, and intents per application class in detail.
KPIs in parentheses in Table \ref{tab:applications} are not inputs for the user experience models, but are part of the evaluation in this paper.

\subsubsection{Remote Terminal Work}

For remote terminal work we define the intent of an administrator typing commands over a Secure Shell (SSH) connection.
An automated SSH client enters commands and measures the duration until the output of the command appears in the terminal.
Only commands which require minimal processing on the server-side, e.g., \textit{uptime} and \textit{date}, are entered.
The SSH connection is established before the start of the experiment.
OpenSSH 7.2 is used as server implementation on Ubuntu 16.04.4 LTS systems.
Client-side automation is implemented using \textit{paramiko}\footnote{http://www.paramiko.org/\label{fn:paramiko}, last accessed: 11.10.2018}.

\subsubsection{File Downloads / Web Browsing}

File download is the batch transfer of a chunk of data over one TCP connection.
As intent we define \textit{emailattach}, a file with random content and a size of \unit[10]{MB}, which is placed on an HTTP server for download.
In an enterprise environment this intent could represent the maximum size of email attachments.
The download is implemented using a short Python script and the \textit{requests} library.
As KPI, the script measures the duration from when the GET request is sent, up to the last received Byte.

Web browsing is implemented using Firefox in version 58.0.2 automated with \textit{selenium}\footnote{https://www.seleniumhq.org/\label{fn:selenium}, last accessed: 11.10.2018}.
The settings are left to the default state and the cache is cleared after every page view.
The number of parallel connections is limited to six per server and HTTP pipelining is not supported anymore by recent Firefox versions.
The connections are configured to be persistent between requests.
The browser interface is disabled (headless mode) and no page rendering is performed in the experiments to minimize the influence of system load and deployed testbed hardware.

This is a scenario where a limited number of browser-based business applications are used frequently and/or all web browsing sessions are tunneled through an enterprise proxy.
With proxies, connections can be persistent even when requesting content from different domains.
General web browsing, where multiple domains are involved without proxy, is not represented well by assuming persistent connections.
This is due to the fact that connection establishment can significantly influence the page load time for longer transport delays.
We define the KPI for one web browsing request as the duration from the initial GET request to the time all embedded resources are received (\textit{page load time}).
For web browsing we define the intent \textit{science\_lab}.
The science\_lab \textsuperscript{\ref{fn:appaware}} template is a web-site with 22 objects with a total size of about \unit[1.3]{MB}.

%We configure the HTTP server to listen on three different ports and send 9 of the object requests to the first port, 8 to the second port and 5 to the third port.
%This mi
%Browsers usually contain objects from multiple HTTP servers and different ports of the same host are treated by the browser as separate HTTP servers.
%Note that this duration does not include the time the browser needs to process the Document Object Model (DOM) and render the content.
%The DOM processing and page rendering steps are CPU-intensive tasks and depend highly on the hardware and system load. 
%We assume this duration to be neglectable compared to the resource transfer duration on dedicated Client PCs in a typical enterprise deployment.

\subsubsection{Adaptive Video Streaming}

HTTP adaptive video streaming is implemented using the TAPAS\cite{de2014tapas} DASH player.
The \textit{conventional} \cite{li2014probe} bit-rate adaptation strategy is selected.
We consider one video view as one request and select the average quality level of all downloaded segments as KPI.
We define the intent \textit{bbb} for on-demand video streaming.
For this intent, we encode the open-source movie Big Buck Bunny in six quality levels with average bit-rates of \kbps{486}, \kbps{944}, \kbps{1389}, \kbps{1847}, \kbps{2291}, and \kbps{2750}.
Only the first \unit[60]{s} of the movie are selected and segmented into 15 chunks of \unit[4]{s} each.
The playback buffer is configured with a maximum size of \unit[60]{s}

Additionally, we define the live-streaming intent \textit{bbb\_live} where the chunk size is reduced to \unit[1]{s} and the buffer is limited to \unit[10]{s}.
Due to encoding overhead for the shorter chunk duration, the bit-rates increase to \kbps{572}, \kbps{1103}, \kbps{1625}, \kbps{2145}, \kbps{2660}, and \kbps{3172}.

\subsubsection{Voice-over-IP}

We emulate VoIP traffic using the Distributed Internet Traffic Generator (D-ITG) by Botta et al. \cite{BottaDP12}.
D-ITG reproduces the inter departure-times and packet sizes of VoIP traffic and measures the KPIs jitter, packet loss, and delay of the resulting UDP packet stream.
We define the intent \textit{G.729.1} for VoIP and configure D-ITG to emulate RTP VoIP calls with the codec G.729.1.
In this configuration, a constant bit-rate stream with 50 packets per second is generated with a packet size of about 20 Bytes ($\approx$ \kbps{8}).

%Next we discuss how we derive the utility value from the KPIs of the applications.

\subsection{Utility from KPIs}
\label{subsec:utilityfromkpis}

\begin{figure}[t]
\centering
\subfigure[Terminal Work ($\UKPI^{\text{(SSH)}}$)]{\includegraphics[width=100pt]{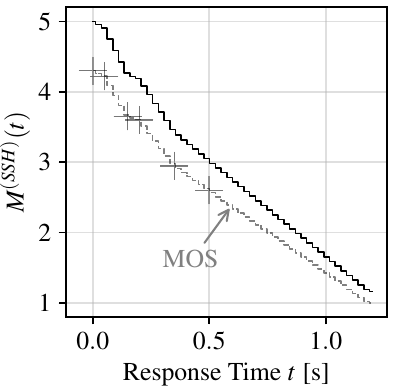}\label{fig:01_ssh_model}}\hspace{25pt}
\subfigure[Web Browsing ($\UKPI^{\text{(WEB)}}$)]{\includegraphics[width=100pt]{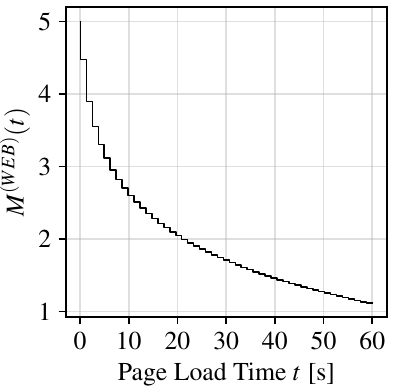}\label{fig:03_web_model}}\hspace{25pt}
\subfigure[File Download (10M) ($\UKPI^{\text{(DL)}}$)]{\includegraphics[width=100pt]{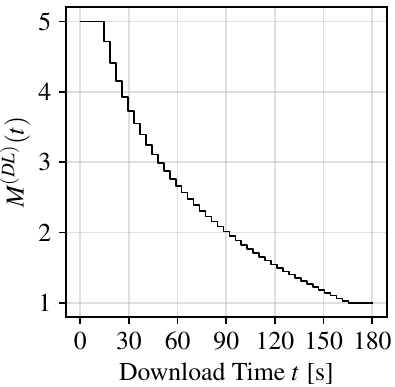}\label{fig:02_webdl_model}}
\vspace{-1em}
\caption{
Utility from application KPIs ($\UKPI$: KPI $\mapsto$ Utility) derived from subjective study results scaled to range $[1,5]$.
$\UKPI^{\text{(SSH)}}$ is derived from subjective study \cite[Fig. 5 (a)]{casas2013quality} by Casas et al. 
Plus signs indicate the MOS data points as collected by the authors in the study.
Web and file download utility values are derived from subjective user studies in \cite{egger2012time} by Egger et al.
}
\label{fig:mos_web_models}
\Description[Utility decreases with response time, page load time and download time.]{Terminal work, web browsing and file download show similar decrease in utility for increased time the user has to wait.}
\end{figure}

We define the current utility value of an application as an estimation of the instantaneous satisfaction of a user with the interaction with the application.
The relationship between KPIs and user experience has to be determined through subjective studies, either directly by conducting dedicated laboratory, field, or crowd-sourcing studies, or indirectly by measuring user-relevant success metrics such as task completion times.
We denote this relationship as $\UKPI$: KPI $\mapsto$ Utility.
In case there is a suitable QoE Mean Opinion Score (MOS) 
%}
model available for the application based on subjective studies, we take a scaled version of the MOS model for $\UKPI$.
Thus, the utility functions are based on the average user experience of the test subjects in the referenced studies.
However, the range of some user experience models does not reach up to 5.0 (Excellent).
In those cases, we define the $\UKPI$ by scaling up the experience model to $[1,5]$.
If no model is available, we define $\UKPI$ based on hand-picked application KPIs.

QoE is an active area of research and holistic models do not exist yet for most applications.
There could be alternatives or more complex models available for the selected user experience models.
Furthermore, custom enterprise applications might require custom user experience studies.
%However, more complex models come at a cost of increased implementation and monitoring effort for the evaluation set-up of this work.
In any case, the presented system design and findings of this paper are independent of the concrete deployed user experience models.
Therefore, the selected models in this work should be seen as rough approximations of the true underlying user experience.

%For an enterprise the options are to create custom models tailored to the deployed applications or to fine-tune existing models from research.
%\fixme{[AB: maybe make it a stronger selling point.]}

\subsubsection{Remote Terminal Work}

We piece-wise interpolate $\UKPI$ for remote typing from the results presented in \cite[Fig. 5(a)]{casas2013quality}.
There, Casas et al. study the QoE of remote desktop services for different use cases.
For the investigated \textit{typing} use case, the test subjects were asked to type a short text on a text processor in a remote desktop session.
The higher the delay in the network, the longer the user has to wait until his actions, e.g., typing a character or deleting character, appear on the screen.
The delay until the actions result in visual feedback is denoted as response time and we choose it as the KPI for remote terminal work.
Figure \ref{fig:01_ssh_model} illustrates the piece-wise interpolated model based on the presented opinion scores in \cite{casas2013quality}.
The authors only investigated response time values up to \unit[0.5]{s}.
We linearly extrapolate the results up to \unit[1.2]{s} where the utility reaches 1.
We define $\UKPI$ as $\UKPI^{\text{(SSH)}}(t) := \MOS^{\text{(SSH)}}(t) - 1) \cdot \frac{4}{3.3} + 1$ to project the MOS values to a utility range of $[1,5]$.

\subsubsection{Web Browsing / File Downloads}

Egger et al. \cite{egger2012time} propose models for the user experience of web browsing and file downloads based on subjective user studies.
The web browsing model uses the page load time as KPI.
For the file download, the download time of a \unit[10]{MB} file is used as KPI.
The MOS value for web browsing is proposed as $\MOS^{\text{(WEB)}}(t) := -0.88 \cdot ln(t) + 4.72$.
For the file download, $\MOS^{\text{(DL)}}(t) := -1.68 \cdot ln(t) + 9.61$.

Figure \ref{fig:03_web_model} illustrates the web browsing model.
The figure highlights the severe impact of the page load time on the user experience in web browsing. 
After only \unit[2.2]{s} waiting time, the MOS is already down from 5 (\textit{Excellent}) to 4 (\textit{Good}).
With additional \unit[4.6]{s} waiting time, the MOS decreases to 3 (\textit{Fair}).
After a total waiting time of \unit[20]{s}, the score ranges between \textit{Poor} and \textit{Bad}.
For web downloads (Figure \ref{fig:02_webdl_model}), the users are more willing to accept longer waiting times.
For example it takes a waiting time of \unit[28]{s} for the opinion score to decrease to 4.
We use the $\MOS^{\text{(DL)}}$ model as proposed by the authors as $\UKPI$ with $\UKPI^{\text{(DL)}}(t) := \MOS^{\text{(DL)}}(t)$.
$\UKPI^{\text{(WEB)}}$ we define as $\UKPI^{\text{(WEB)}}(t) := (\MOS^{\text{(WEB)}}(t) - 1) \cdot \frac{4}{3.6} + 1$.

\subsubsection{Adaptive Video Streaming}

The user experience during an adaptive video streaming session depends on factors such as average presented quality, number and amplitude of quality switches, frequency and duration of stalling events, device's screen size, viewing environment, user expectation, encoding, adaptation strategy, and content type \cite{hossfeld14assessingeffect}.
To the best of our knowledge there is no holistic model for the user experience of adaptive streaming available at the moment.
One option for enterprises is to create custom models, for example for onboarding videos for new employees.

Studies show the average quality as a dominant influence factor \cite{hossfeld2015identifyingqoeoptimal} for the user QoE.
We therefore assign a utility value to a streaming application based on the observed average quality $q^{\text{(avg)}}$ and the maximum and minimum quality level, $q^{\text{(max)}}$ and $q^{\text{(min)}}$.
The utility value is then determined by $\UKPI^{\text{(HAS)}}(q^{\text{(avg)}}) := \frac{q^{\text{(avg)}} - q^{\text{(min)}}}{q^{\text{(max)}} - q^{\text{(min)}}} \cdot 4 + 1$.

\subsubsection{Voice-over-IP}

Sun et al. \cite{sun2006voice} propose a model for the MOS of VoIP depending on the used audio codec and a user's interactivity, i.e., whether the user is only listening or also conferencing.
The MOS value is presented as polynomial equation with constants $a$ to $j$ and with packet loss ratio and delay as input parameters.
The constants depend on the used codec.
We configure D-ITG to emulate G.729. 
The MOS model $\MOS^{\text{(VoIP)}}(\text{loss}, \text{delay})$ is then described by Eq.~10 and Table~II in \cite{sun2006voice}.
We define the $\UKPI$ accordingly as $\UKPI^{\text{(VoIP)}}(\text{loss}, \text{delay}) := \MOS^{\text{(VoIP)}}(\text{loss}, \text{delay}) - 1) \cdot \frac{4}{2.65} + 1$.

%(Eq. \ref{eq:voip_poly}).
%\vspace{-1.5em}
%\begin{equation}
%\label{eq:voip_poly}
%\begin{split}
%MOS_{VoIP}(loss, delay)=3.61-0.13 \cdot loss+1.22 \cdot 10^{-3} \cdot delay+3.76 \cdot 10^{-3} \cdot loss^2-2.29 \cdot 10^{-5} \cdot delay^2 + \\ 
%4.71 \cdot 10^{-6}\cdot loss \cdot delay-5.16 \cdot 10^{-5} \cdot loss^3+2.54 \cdot 10^{-8} \cdot delay^3 +\\
%1.28 \cdot 10^{-7} \cdot loss \cdot delay^2-4.43 \cdot 10^{-8} \cdot loss^2 \cdot delay
%\end{split}
%\end{equation}

\subsection{Utility Functions}

\begin{figure}[t]
\centering
\includegraphics[width=390pt]{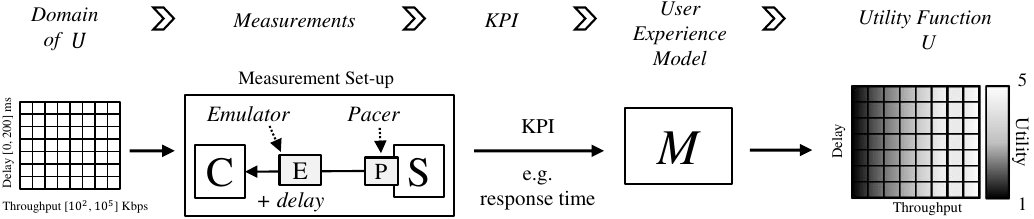}
\caption{
Utility functions for \textit{(class, intent)} are generated by first defining a measurement domain in terms of throughput and delay.
Second, the domain is quantized and the application KPIs are measured in an emulated network environment using the quantized parameters for throughput and delay.
Third, user experience models are used to derive the utility for the measured parameters.
}
\label{fig:util_testbed}
\Description[A workflow from defined parameter space to measured utility function.]{Domain of U as input for measurement setup, which outputs application KPI to the user experience model which is the input for the utility function.}
\end{figure}

The utility function $\Ux_{a}$: (Throughput \text{[Kbps]}, Delay \text{[ms]}) $\mapsto [1, 5]$ approximates the QoE-aware utility for a specific application type $a$ for a unidirectional pacing rates and maximum delay threshold using the utility model.
Hence, the function solves the problem of linking network resource demands with the resulting user experience.
The hereinafter described methodology for constructing the utility functions can be applied in an automated fashion to any enterprise application and its intents.

Figure~\ref{fig:util_testbed} illustrates the process of constructing the utility functions.
A set-up measures the utility of each application and intent for different pacing rates and delays in an isolated environment.
Two hosts (Host S and Host C) are connected through a network emulator.
On the emulator, Linux \textit{netem} is adding delay to all packets passing through it.
Host S is running the server endpoint of the application, e.g., in case of web browsing an HTTP web server.
The client endpoint is assigned to Host C, e.g. the web browser.
Host S egress traffic is paced using the \textit{cfg} queuing discipline (Section~\ref{subsec:pacingimpl}).
From the measurements we derive the 2-dimensional utility functions.
Note that to account for asymmetric data-rates in a conversation, which is the case for the most server-client traffic such as web traffic, the two directions of a conversation have to be described by different utility functions.
For the sake of simplicity, we consider only one direction per conversation as constrained and only present the server-to-client utility functions.
%Server to client traffic is prevalent for end-user applications , file download and video streaming.
For the throughput, we measure \dl{} in the range of $[100, 5000]$~Kbps, \web{} in the range of $[100, 12000]$~Kbps, \vod{} and \live{} in the range of $[750,5000]$~Kbps and \voip{} and \ssh{} in the range of $[100, 500]$~Kbps.
For the delay, we measure \web{}, \dl{}, \vod{}, \live{}{} in the range of $[0, 240]$~ms and \voip{} and \ssh{} in the range of $[0, 500]$~ms.
The maximum pacing rate per intent is set so that further increasing the pacing rate does not improve the utility for any delay demand.
%The measurement ranges also restrict the output domains of the resulting utility functions.

%\begin{figure}[b]
%\centering
%\includegraphics[width=240pt]{figures/util_testbed.pdf}
%\label{fig:util_testbed}
%\caption{
%Testbed to measure the utility of applications for different pacing rates and delays. Two hosts connected through a bridge which adds delay.
%}
%\end{figure}

%
%\begin{figure}[h]
%\centering
%\subfigure[File Download]{\includegraphics[width=0.442\columnwidth]{figures/06_util_grid_webdl.pdf}\label{fig:02_webdl_model}}
%\subfigure[Web Browsing]{\includegraphics[width=0.47\columnwidth]{figures/06_util_grid_web.pdf}\label{fig:03_web_model}}\\
%\caption{.}
%\label{fig:util_grids_web}
%\end{figure}
%
%\begin{figure}[h]
%\centering
%\subfigure[DASH]{\includegraphics[width=0.46\columnwidth]{figures/06_util_grid_dash.pdf}\label{fig:02_webdl_model}}
%\subfigure[SSH]{\includegraphics[width=0.47\columnwidth]{figures/06_util_grid_ssh.pdf}\label{fig:03_web_model}}\\
%\caption{.}
%\label{fig:util_grids_ssh_dash}
%\end{figure} 

\begin{figure}[t]
\centering
\subfigure[\dl{}]{\includegraphics[width=105pt]{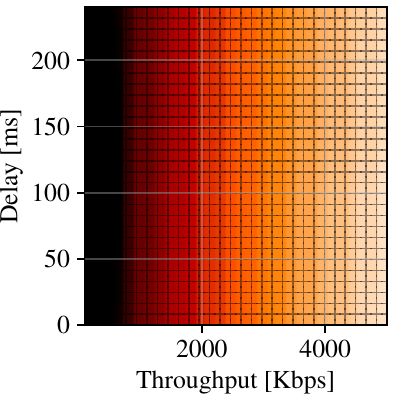}\label{fig:06_util_grid_webdl}}
\subfigure[\web{}]{\includegraphics[width=105pt]{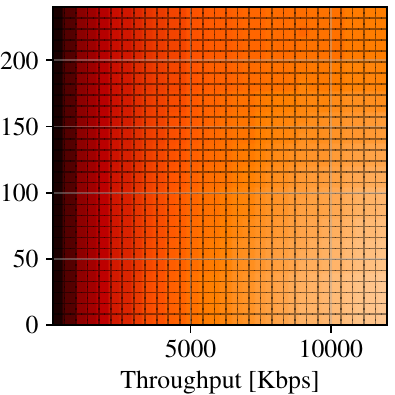}\label{fig:06_util_grid_web}}
\subfigure[\vod{}]{\includegraphics[width=105pt]{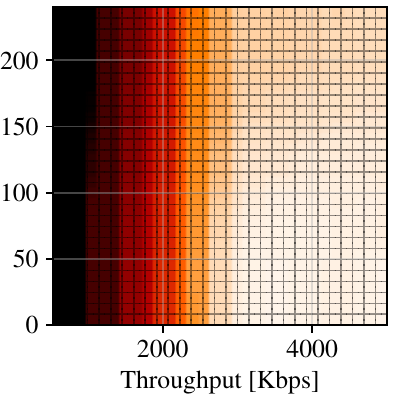}\label{fig:06_util_grid_dash}}
\subfigure[\live{}]{\includegraphics[width=105pt]{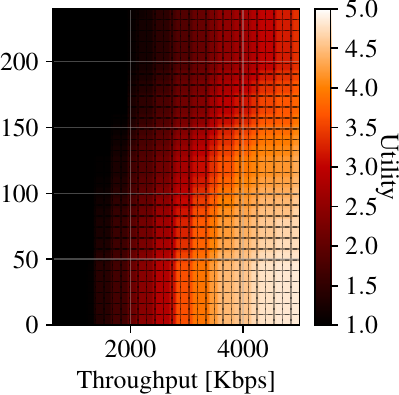}\label{fig:06_util_grid_dash_live}}\\
%\subfigure[SSH (\textit{sshadmin})]{\includegraphics[width=0.47\columnwidth]{figures/06_util_grid_ssh.pdf}\label{fig:03_web_model}}\\
\vspace{-1em}
\caption{
Utility functions $\Ux_{a}$: (Throughput, Delay) $\mapsto [1, 5]$ which map throughput and delay to utility for the application classes file download, web browsing and video streaming and intents defined in \reftab{tab:applications}.
}
\label{fig:util_grids}
\Description[Heatmap of throughput and delay values to utility.]{Heatmap with decreasing utility for increasing throughput and decreasing delay. Live streaming decreases fastest.}
\end{figure}

Figure \ref{fig:util_grids} presents the measurement results for the utility of the applications depending on delay and throughput.
The intersections of the grid indicate the quantization as used by the resource allocation problem formulation.
The figure shows that \dl{}, \web{}, and \vod{} are highly dependent on the throughput and only a minor dependency on delay is visible.
\live{} depends on delay and throughput.
\ssh{} (not shown) depends solely on the delay.
For \dl{} (Fig.~\ref{fig:06_util_grid_webdl}), the impact of the delay is limited to the TCP handshake, the file request and acknowledgements packets.
The impact is insignificant compared to the download time and not visible on the figure.
For \vod{}, the impact of delay depends additionally on the number and playtime duration of video segments and the adaptation strategy. 
As illustrated by Figure~\ref{fig:06_util_grid_dash}, the influence of delay for the intent \vod{} is minor. 
For \live{} there is a clear influence of delay on the utility (Fig.~\ref{fig:06_util_grid_dash_live}).
For \ssh{}, the delay is the important influence factor, as every typed character triggers an outgoing packet and requires an immediate response packet.
As we use persistent HTTP connections for web browsing, there is no influence of the delay on the \web{} utility due to the TCP handshake.
The influence of the delay is limited to the requests of the HTML index object and the embedded resources (Fig.~\ref{fig:06_util_grid_web}).

The maximum utility values an application can reach in the measurements are determined by implementation-specific factors and the domain and range of the utility function.
For example \web{} is limited by the browser processing time and \vod{}/\live{} depend on the behavior of the adaptation algorithm.
\ssh{} can reach the highest utility of 5 with \kbps{100} throughput and \unit[0]{ms} delay.
\voip{} can reach 5 with \kbps{100.0} and \ms{34.5}.
For \web{} the highest utility is 4.5 with \kbps{11589.7} and \ms{33.1} delay.
\vod{} can reach its highest utility of 4.9 with \kbps{3479.3} and \ms{41.4}.
\live{} can reach 4.8 with \kbps{5000.0} and \ms{41.4}.
\dl{} can reach a utility of 4.8 with \kbps{5000} and \ms{99.3} delay. 

%max - web: 4.5 at 11589.7 Kbps [100.0 to 12000.0] with 33.1 ms [0.0, 240.0] delay
%max - webdl: 4.8 at 5000.0 Kbps [100.0 to 5000.0] with 99.3 ms [0.0, 240.0] delay
%max - ssh: 5.0 at 100.0 Kbps [100.0 to 500.0] with 0.0 ms [0.0, 500.0] delay
%max - video: 4.9 at 3479.3 Kbps [100.0 to 5000.0] with 41.4 ms [0.0, 240.0] delay
%max - dash_live: 4.8 at 5000.0 Kbps [100.0 to 5000.0] with 41.4 ms [0.0, 240.0] delay
%max - voip: 5.0 at 100.0 Kbps [100.0 to 500.0] with 34.5 ms [0.0, 500.0] delay

%% file: content/06_networkwide_app_fairness.tex
%\section{Network-wide Application Fairness Formulation}
%\label{sec:fairness}

%In the subsequent section we discuss the experiment design and evaluation methodology.

\section{Utility Allocation Problem} \label{subsec:problem_formulation}

%We define fair shares in terms of the utility of an application.
The network controller performs the calculation of the shares to be allocated based on the number of applications, their utility functions, the network topology, current network status, and fairness criteria.
For this paper we define the utility fairness criteria as follows.
We first try to maximize the minimum utility over all applications (\textit{max-min-fairness}) and afterwards maximize the sum of utilities while allowing a small decrease in minimum utility.
The complete allocation formulation is introduced in Appendix \ref{appendix:problem}.

We formulate the problem as a Mixed Integer Linear Program (MILP). 
The objective of the MILP in the first step is to maximize the minimal utility value $\UVMIN$ over all applications.
In the second step the MILP maximizes the sum of all utility values, while the minimum utility $\UVMINII$ is restricted to the range $\UVMINII \in [(\UVMIN1 - \epsilon), \UVMINI]$ with $\epsilon = 0.3$.
The MILP has to consider the two-dimensional utility function of every application, the capacities of all paths between application endpoints, and the delay at intermediate hops depending on the link utilization.
The decision variables describe which pacing rate to apply to which application and how to configure the routing between application endpoints.

We allocate a specific data-rate per application.
Hence, we do not consider how much data-rate is actually consumed by an application.
On the one side, static allocation via application pacing can guarantee predictable application performance as this work shows in the experiments.
But on the other side, there is no statistical multiplex gain in case the applications use less resources than allocated to them. 
As a consequence, the network may be under-provisioned and available resources are potentially not made available to other applications.

In this work we configure all applications in the experiments to constantly use the link which makes the number of active applications equal to the total number of applications on a given link. 
That puts the most stress on the link for a given number and mix of applications. 
Reducing the activity of an application would be equivalent to reducing the number of simultaneously active applications on the link.
But existing research on Internet traffic and congestion can be leveraged by future work, e.g., to overprovision the links based on the actual number of active applications at a given point in time.

\subsection{First Step: Maximize Minimum Utility}

$\mathcal{A}, a \in \mathcal{A}$ is the set of all unidirectional application flows $a$.
We define $\TU(a)$ as the target utility value of an application flow $a$.
In the first step we maximize the minimum utility value (\textit{max-min fairness}) subject to all application utilities have to be larger than the minimum utility value $\UVMIN$. 

\vspace{-0.5em}
%\begin{equation}
\begin{align*}
\text{maximize:} & \quad \UVMIN \\
\text{subject to:} & \quad \TU(a) \geq \UVMIN  \quad \forall a \in \Ax \\
                   & \quad \text{and (\ref{eq:a_tp}) - (\ref{eq:last_delay_constr}) in appendix A2 - A6.}
\end{align*}
%\end{equation}

\noindent
We denote the optimal value of $\UVMIN$ of the first step as $\UVMINI$.
A full definition of all symbols is provided in the appendix A.

\subsection{Second Step: Maximize Sum of Utilities For Constrained Minimum Utility}

In the second step we relax the max-min constraint by $\epsilon$ and maximize the sum of all target utility values.
We add the additional constraint to bound $\UVMIN$ by $\UVMINI - \epsilon = 0.3$:

\vspace{-0.5em}
%\begin{equation}
\begin{align*}
\text{maximize:} & \quad \sum_{a \in \mathcal{A}} \TU(a) \\
\text{subject to:} & \quad \UVMIN \geq \UVMINI - \epsilon \\
                   & \quad \text{and (\ref{eq:a_tp}) - (\ref{eq:last_delay_constr}) in appendix A2 - A6.}
\end{align*}
%\end{equation}
\vspace{0.05em}

\noindent
For the remainder of the paper, if not otherwise stated, $\UVMIN$ denotes the optimal value of the second step ($\UVMINII$).
The complete formulation of the problem can be found in Appendix~\ref{appendix:problem}.

%% file: content/07_eval_methodology.tex
\section{Experiment Design and Set-up} \label{sec:evaluation}

The objective of the experiments is to show the dependability and scalability of resource allocation via end-host pacing and how the different application classes profit and/or suffer from the enforced packet pacing.
The experiments are conducted in a set-up where we monitor sets of increasing number of parallel applications sharing a throughput-constrained link.
For each set of applications we measure the utility with and without resource allocation and discuss the differences in the evaluation.
Dynamic embedding of applications at run-time and additional intents are out of scope of this evaluation.
Next, we elaborate on the deployed experimental set-up (Section \ref{subsec:expsetup}) and the custom pacing implementation (Section \ref{subsec:pacingimpl}).
Afterwards, we discuss the experiment parameters (Section \ref{subsec:paraspace}).
The results of the evaluation are presented in the subsequent Section \ref{sec:results}.

\subsection{Experiment Set-up} \label{subsec:expsetup}
\begin{figure}[t]
\centering
\includegraphics[width=410pt]{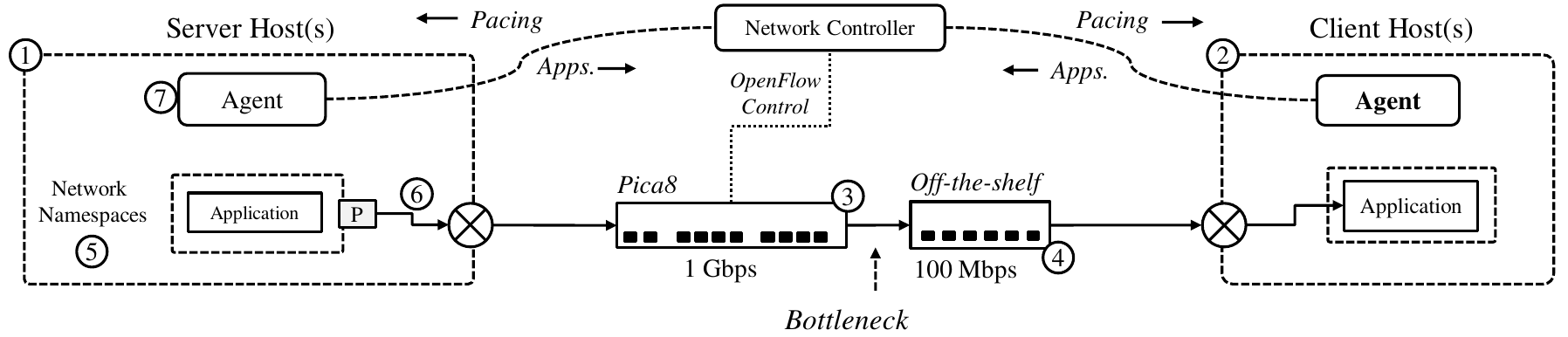}
\vspace{-1em}
\caption{
Experimental set-up.
Two groups of hosts, one server and one client group, are connected via an SDN-capable switch and an unmanaged \mbps{100} link to each other.
A network controller calculates fair shares, configures the pacers, and collects statistics.
}
\label{fig:09_proc}
\Description[Network controller configuring agents on server and client hosts and switches.]{Pacing is applied by controller to agents on servers and clients and to switches. A bottleneck between a Pica8 and off-the-self 100 Mbps switch.}
\end{figure}

Figure \ref{fig:09_proc} illustrates the experiment set-up, consisting of two groups of hosts: one server \fkey{1} and one client group \fkey{2}.
The link between the two groups is throughput-constrained and the applications running on the host groups have to share the limited bandwidth.
The network consists of two switches, one SDN-enabled Pica8 P-3290 \fkey{3} and one unmanaged off-the-shelf \mbps{100} switch \fkey{4}. 
The link between the two switches constrains the available data-rate between the hosts on the left and on the right side to \mbps{100}.
The Pica8 switch is equipped with a maximum queue size of \unit[1]{MB} and maximum queuing delay of about \ms{80} towards the \mbps{100} link.
We deploy three modern desktop PCs on each side to meet the processing and memory resources required by the experiment scenarios.

Each application consists of a server and client endpoint, e.g., a web server and a browser.
All endpoints are confined to a separate network namespace \fkey{5} and connected via virtual interfaces and a software bridge to the host's physical interface \fkey{6}.
Each namespace is configured with a unique IP and MAC address.
Furthermore, every client is connected to an exclusive server application.
That way, the pacing rate can be set per namespace and no further control is needed to assign outgoing server packets to different pacers.
In case of web browsing, video streaming, and web download, each client is assigned to an exclusive light-weight HTTP server, but with shared content.
The server endpoints are placed left of the bottleneck and the client endpoints to the right of the bottleneck, which makes the egress queue and interface of the Pica8 the bottleneck.
Pacers (\pacer{}) based on our \textit{cfq} implementation (Section~\ref{subsec:pacingimpl}) restrict the egress rate of the namespaces/applications towards the hosts' software bridges.

%The queuing time at the pacers and the switch is measured by a custom in-band RTT measurement tool based on small UDP packets (not shown in Figure \ref{fig:09_proc}).
All management and monitoring operations are performed out-of-band.
The KPIs of each application are measured at the client endpoint, e.g., the page load time at the browser, and reported to the network controller by the applications' agents \fkey{7}.
Additionally, we frequently poll the statistics counters of all physical and virtual network interfaces to measure throughput, queue length and packet loss.

\subsection{Pacing Implementation}\label{subsec:pacingimpl}

%A pacing implementation on the end-hosts eliminates packet bursts and ensures rate conformance.
In Linux, pacing is implemented as a \textit{queuing discipline}.
Furthermore, a mechanism called TCP small queues\footnote{\url{https://lwn.net/Articles/506237/}, Accessed: 2018-10-12} exerts backpressure on the applications to mitigate buffer bloat and packet loss by limiting the allowed number of Bytes per flow in the queuing discipline and device queue (default: \unit[128]{Kilobytes}).
Other operation systems offer similar pacing mechanisms.
%Application pacing is commonly implemented as a part of the networking stack of a device's operating system.
%Next we give details on pacing in the Linux kernel and the modifications required for our purpose.
%Linux-based systems commonly deployed on servers, smartphones and partly on desktop systems.
We implemented a custom queuing discipline based on the existing Fair Queuing (\textit{fq}) discipline \footnote{\url{https://lwn.net/Articles/564825/}, Accessed: 2018-10-12}, referred to as Custom Fair Queuing (\textit{cfq}).
Every conversation defined by \textit{(class, intent)} and by one or multiple sockets, can be assigned to an exclusive queue with a target packet release rate as configured by the network controller through the local agent.
Packets from the queues are released time-based.
The departure time of the next packet $time\_next\_packet$ is determined by the current time $now$, the size of the current packet $pkt\_len$ and the target pacing rate $target\_rate$: $time\_next\_packet = now + \frac{pkt\_len}{target\_rate}$
%Details on the implementation are available in the released source code.

\subsection{Parameter Space and Experiment Procedure}\label{subsec:paraspace}

The parameter space of the experiments is limited to the \textit{number} and \textit{types} of the applications and whether the experiment is \textit{managed} or \textit{best effort}.
In detail, the bottleneck link is shared by $\{2,4,..,24\}$ applications per class, in total $|\Ax| \in \{10, .., 120\}$.
For video streaming, half of the applications are of type \live{} and the other half of \vod{}.

At the start of the experiment, a configuration file is pushed to each host telling the host which number and type of applications to start.
Each application is modified to start in its own network namespace and to report its type to the host-local agent (\protect\fkeynb{5} in Fig.~\ref{fig:09_proc}).
The local agent forwards this information to the network controller.
Once all applications are registered with the network controller, the controller calculates the resource shares of utility for each application and pushes the corresponding static pacing rates to the agents.
The agents configure the pacers of the applications' network namespaces accordingly.
The pacing rate is not changed during an experiment run.
The SDN-enabled Pica8 switch is configured via OpenFlow for simple forwarding.
Besides the forwarding rule configuration, the OpenFlow connection is used to poll queue and interface statistics.

The duration of one experiment run is \unit[15]{minutes} with an additional \unit[1]{minute} warm-up and cool-down phase.
The applications are started at random times during the warm-up phase and requests during the warm-up or cool-down phase are discarded for the evaluation.
Each experiment is repeated 11 times.
If an application's request is finished, it initiates a new request after a pause time of \unit[100]{ms}.
One request equals one video view for \vod{} and \live{}.
For \voip{}, one request equals one \unit[30]{s} phone call.
The reason for the static pause time of \unit[100]{ms} is that this results in an almost constant number of concurrent applications using the bottleneck link. 
Hence, each application in a specific scenario is constantly sending/receiving requests/responses, except of a \unit[100]{ms} break between requests to allow for a reset of an application's state. 
Increasing the pause time between requests would effectively decrease the number of concurrently active applications at a specific point in time.
\textit{Cubic} is configured as TCP congestion control algorithm.
Cubis is chosen as comparison as it shows better performance on congested links compared to Compound and New Reno TCP \cite{abdeljaouad2010performance} and it is the default algorithm for many Linux server variants.
BBR congestion control proposed by Google fails to show performance benefits and fairness in heterogeneous environments \cite{hock2017experimental} compared to Cubic.

There exist valid optimal solutions to the allocation problem formulation with applications of the same type to be assigned different utility values.
For easier presentation of the results, we constrain the problem formulation to choose one utility value per type.
The bottleneck link is modeled with a capacity of \mbps{100}.
%The increased capacity (\mbps{100} assumed capacity vs. approximately \mbps{100} actual capacity) compensates for the over-provisioning through fixed rate allocation.
As the sum of all paced flow rates does not exceed the available capacity, and due to the short pause times between application requests, the link in the managed case is slightly under-provisioned.
Thus, a large queue build-up is unlikely and the link delay of the bottleneck is modeled with a constant delay of \ms{2}.
In the best effort case, the link is already over-utilized with 10 competing applications and experiences \unit[58]{ms} delay and \unit[0.5]{\%} packet loss (discussed later in Section~\ref{subsec:voip}).

We provide details on how the experiment setup is expressed in the terms of the variables of the theoretical problem formulation in Appendix \ref{appendix:valuesetup}.

%% file: content/08_results.tex
\section{Evaluation} \label{sec:results}

We evaluate the performance of an increasing number of applications sharing a throughput-constrained link with and without data-rate management.
The evaluation is pursuing the following questions.
\begin{enumerate*}[label=\roman*)]
\item How does the minimum and average utility of the applications compare between the managed and best effort scenarios?
\item Which applications benefit, which utility values are decreased, and why?
\item Can pacing result in configurable and thus predictable application performance in terms of the difference between the target and the measured utility?
\item How fair, in terms of utility, are the best effort and the managed utility distribution?
\end{enumerate*}

First, we evaluate how the available data-rate is distributed among the applications in a best effort scenario and present the resulting utility distribution.
Second, we solve the allocation formulation for the scenario, implement static pacing in the set-up for each application and present the gains in terms of utility.
Third, we present how pacing affects the QoS parameters, such as packet loss and jitter, of the link.
Fourth, we conduct a parameter study on the number of parallel applications and show how the gains and fairness changes with increasing number of parallel applications.
Error bars in the result figures indicate the standard deviation if not otherwise stated.
In cases the error bars are not clearly visible on the presented scale, they are omitted from the figures.

\subsection{Best Effort Throughput and Utility Distribution}

\begin{figure}[t]
\centering
\subfigure[Throughput]{\includegraphics[width=180pt]{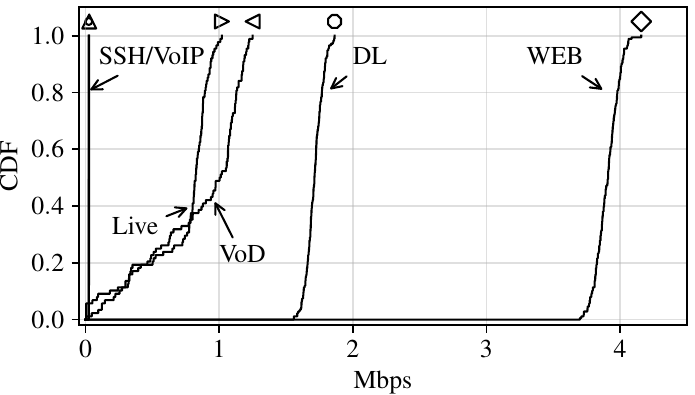}\label{fig:25_app_traffic}} \hspace{25pt}
\subfigure[Utility]{\includegraphics[width=185pt]{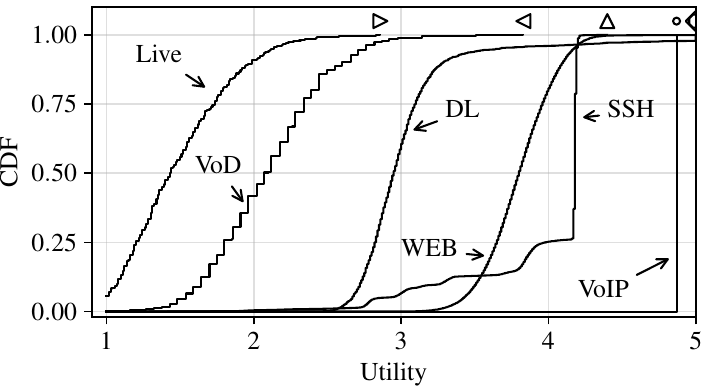}\label{fig:27_a80_unpaced_cdf}}
\vspace{-10pt}
\caption{
Best effort throughput and utility of the different application types for 16 clients per application class.
The markers at the top are for better visual indication of the application types.
}
\label{fig:best_effort_A_80}
\Description[Throughput and utility is unfairly distributed.]{Distribution of throughput and utility shows web browsing, VoIP and live streaming doing exceptional well and bad, respectively.}
\end{figure} 

First, we take a close look at the best effort application performance for single scenario with 16 clients per application class, $16 \times \web{}$, $16 \times \dl{}$, $16 \times \ssh{}$, $16 \times \voip{}$, $8 \times \vod{}$, $8 \times \live{}$, in total $|\Ax| = 80$.
The scenario with 80 applications is selected for a closer inspection due to the fact that among the investigated application counts (from 10 up to 120 applications), one of the highest gains is observed here.
With the 80 applications competing for the bandwidth, the link is fully utilized resulting in an average packet loss of \unit[4]{\%} and queuing delay of \unit[80]{ms}.

Figure~\ref{fig:25_app_traffic} presents the CDFs of the average throughput and Figure~\ref{fig:27_a80_unpaced_cdf} the CDFs of the utility values of all requests per application type.
Multiple observations can be made from the figures.
First, the throughput as well as the utility is distributed non-uniformly between the application types.
For example, while \web{} enjoys high throughput and utility (median $\geq \mbps{3.9}$, 3.8 utility), \live{}'s achieved throughput is less than \mbps{1} and median utility is about 1.4.
\web{}'s high throughput is due to the use of multiple persistent parallel TCP connections,
while video streaming clients, \dl{}, and \ssh{} establish only one TCP connection.
Parallel TCP connections allow an application to receive a proportional larger fraction of the available throughput.
As web download has no idle periods during the download, web download exhibits a higher average throughput than video streaming.

Second, even \vod{} and \live{}, which belong to the same application class (video streaming) and achieve similar throughput rates, suffer from unfair utility distribution (1.3 vs. 2.1).
This is due to the smaller playback buffer for live streaming and the increased encoding overhead for the shorter video chunks. 
%Details on the video streaming performance is given in Section \ref{subsec:dash}.
Third, the average throughput of \ssh{} and Voice-over-IP (\voip{}) is below \kbps{100}, while the utility is 3.7 and 4.9, respectively.
\ssh{}'s performance is influenced by delay, caused by queuing at the bottleneck link, and retransmissions, due to lost packets when the bottleneck's queue is overflowing.
\voip{} is barely influenced in this scenario, as the maximum delay and packet loss over the single bottleneck is acceptable for \voip{} traffic according to the user experience model.
Details on the performance of \voip{} is given in Section \ref{subsec:voip}.
Fourth, the utility distributions per application type are varying with a standard deviation of 0.2 (\web{}) to 0.5 (\dl{}), with the exception of \voip{}.
Hence, application performance is not consistent across requests of the same application type, and, as a consequence, there is an unfair distribution of shares, even within the same application type.

% == unpaced ==
% median(model_web) = 3.8
% std(model_web) = 0.2
% median(model_webdl) = 2.8
% std(model_webdl) = 0.5
% median(model_video) = 2.3
% std(model_video) = 0.3
% median(model_video_live) = 1.3
% std(model_video_live) = 0.3
% median(model_ssh) = 3.7
% std(model_ssh) = 0.4
% median(model_voip) = 4.9
% std(model_voip) = 0.0

%median(model_web) = 4.1 Mbps
%median(model_webdl) = 1.6 Mbps
%median(model_video) = 1.0 Mbps
%median(model_video_live) = 0.7 Mbps
%median(model_ssh) = 0.0 Mbps
%median(model_voip) = 0.0 Mbps

In summary, best effort delivery is inadequate to provide fair and consistent application performance for multiple applications sharing a constrained link.
Best effort delivery does not consider different demands (throughput vs. delay-sensitivity), transport protocols (TCP vs. UDP), or multiple flows per application.
Furthermore, the constrained link is overloaded, resulting in lost packets and queuing delay.

\subsection{Managed Utility Distribution}

\begin{figure}[t]
\centering
\subfigure[\vod{}]{\includegraphics[width=100pt]{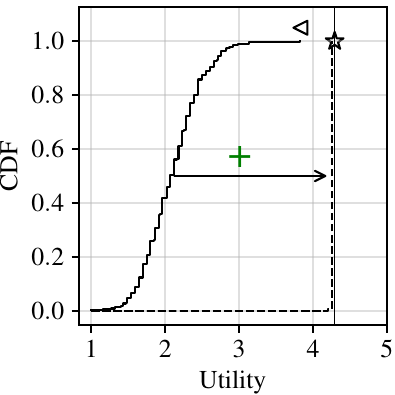}\label{fig:28_a80_gain_cdfs_model_video}}
\subfigure[\live{}]{\includegraphics[width=100pt]{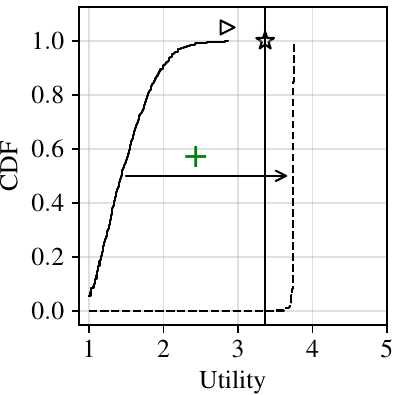}\label{fig:28_a80_gain_cdfs_model_video_live}}
\subfigure[\ssh{}]{\includegraphics[width=100pt]{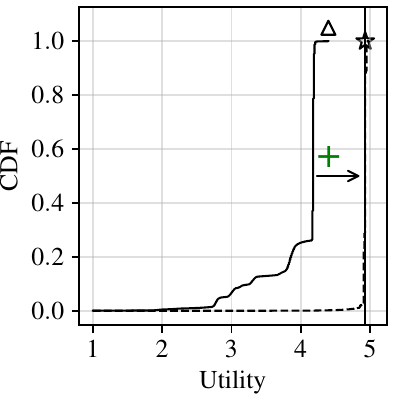}\label{fig:28_a80_gain_cdfs_model_ssh}}
\subfigure[\web{}]{\includegraphics[width=100pt]{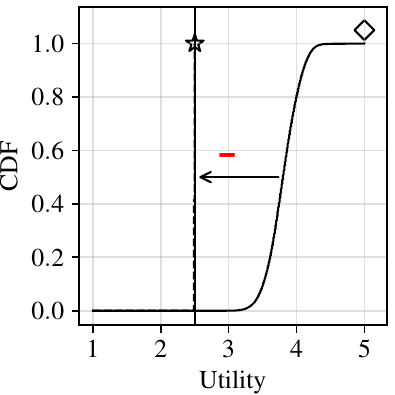}\label{fig:28_a80_gain_cdfs_model_web}}\\
\subfigure[\dl{}]{\includegraphics[width=100pt]{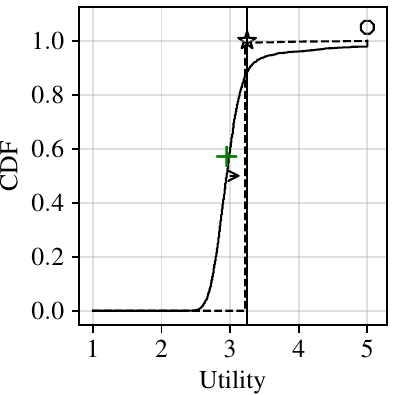}\label{fig:28_a80_gain_cdfs_model_webdl}}
\subfigure[\voip{}]{\includegraphics[width=100pt]{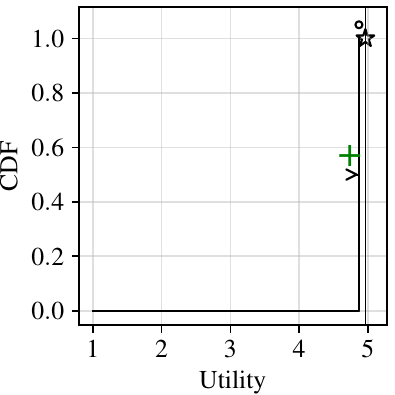}\label{fig:28_a80_gain_cdfs_model_voip}} \hspace{10pt}
\subfigure[Best effort standard deviation]{\includegraphics[width=157pt]{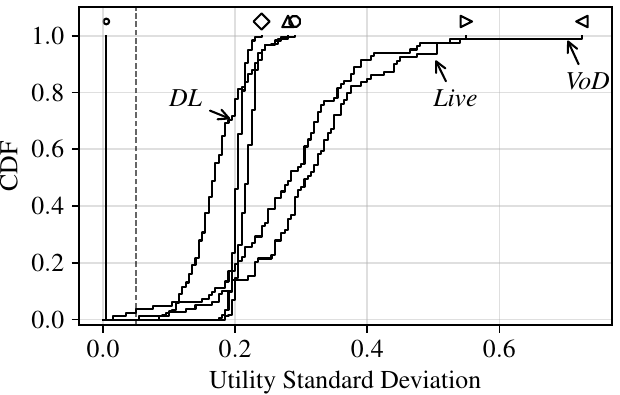}\label{fig:30_a80_std_cdf_unpaced}}
%\subfigure[std managed]{\includegraphics[width=100pt]{figures/30_a80_std_cdf_paced.pdf}\label{fig:20a_per_app_std_paced}}
\vspace{-1.2em}
\caption{
Figures \subref{fig:28_a80_gain_cdfs_model_video} to \subref{fig:28_a80_gain_cdfs_model_voip} show measured best effort and managed application utility for $|\Ax| = 80$ applications sharing the constrained link.
The dashed lines indicate the utility CDF for the managed scenario, the solid lines the best effort scenario.
The star and vertical line mark the target utility $\TU$ for the application type.
Arrows to the right highlight the improvement in median utility.
Figure \subref{fig:30_a80_std_cdf_unpaced} show the standard deviations of a client's utility values per application type.
}
\label{fig:28_a80_gain_cdfs}
\Description[Decrease and increase in average utility distribution.]{CDFs of applications' increase and decrease in utility by management compared to best effort.}
\end{figure} 

Next, we solve the allocation problem formulation with the max-min fairness criteria for the scenario with 80 parallel applications and apply the calculated pacing rates.
Figures \ref{fig:28_a80_gain_cdfs_model_video} to \ref{fig:28_a80_gain_cdfs_model_voip} illustrate the best effort (solid lines) and managed utility (dashed lines) for the scenario with 16 clients per application class.
Improvements in median utility due to the data-rate management are indicated by ($\rightarrow$, $+$).
Deteriorations are shown by ($\leftarrow$, $-$).
The target utility per application type, as calculated by the allocation formulation, is indicated by ($|$, $\star$).

The figures \subref{fig:28_a80_gain_cdfs_model_video} to \subref{fig:28_a80_gain_cdfs_model_voip} show that all application types, except \web{} and \voip{}, profit from the management.
\live{} benefits most from the management, with a median increase of 3.1 (from 1.3 to 4.4).
\vod{}, \ssh{} and \dl{}'s median utility improve by 2.0, 1.0, and 0.4, respectively.
On the other hand, \web{}'s median utility decreases by 1.3 (from 3.8 to 2.5).
With pacing \web{} can not get an unfair advantage over the other applications by using multiple parallel TCP connections. 
No noteworthy improvement or deterioration in utility is measurable for \voip{}.

\live{} \subref{fig:28_a80_gain_cdfs_model_video_live} exhibits a deviation of about 0.5 between the target and measured utility.
The deviation is the result of an inaccuracy in the live streaming utility function.
The samples collected from the utility measurement setup are supplemented with interpolated values to build the quantized utility function.
In the case of live streaming and low delay values, the interpolation results in a utility error of about 0.5.
The error can be reduced by collecting more measurement samples from the throughput-delay parameter space and/or fine-tuning the interpolation algorithm.

Figure~\ref{fig:30_a80_std_cdf_unpaced} presents the \textit{standard deviations per client} of a specific type for the \textit{best effort} scenario.
The smaller the standard deviation is, the more consistent is the experience of a single user.
The dashed vertical line indicates the maximum ($=0.05$) of the standard deviations in the managed case (per type CDFs are not shown for the managed case).
\dlwm{} clients exhibit the largest median standard variation (0.64) among the application types, followed by \sshwm{} with 0.41.
\webwm{} clients' median variation is the second smallest with 0.25.
There is no visible variation for \voipwm{}.
The figure also shows that not only the utility value per client request varies, but also the behavior of each client.
For example for \vodwm{}, the standard deviation varies between 0.1 and 0.43.
Hence, some clients experience a smaller quality variation for their video views than other clients.

\subsection{Link QoS and VoIP Performance Details} \label{subsec:voip}

Next, we take a closer look at the QoS metrics of the constrained link in terms of packet loss, queuing delay and jitter for an increasing number of parallel applications.
In the best effort case we expect the link QoS parameters to degrade because the link is fully saturated and the interface queue is overflowing. 
In the managed case we do not expect any degradation as the level of link saturation is managed.
As the MOS and utility functions of \voip{} are based on the QoS metrics, we also discuss why the QoS metrics have only minor influence on the \voip{} performance in the evaluation.

\begin{figure}[t]
\centering
\subfigure[Packet Loss]{\includegraphics[width=106pt]{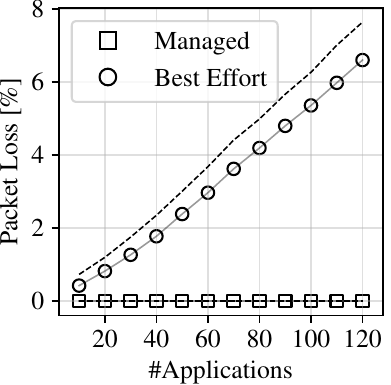}\label{fig:22_voip_metrics_loss}} \hspace{15pt}
\subfigure[RTT]{\includegraphics[width=106pt]{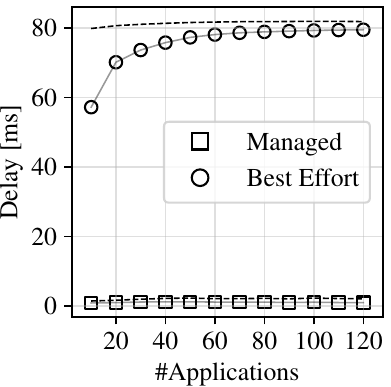}\label{fig:24_delay_delay}} \hspace{15pt}
\subfigure[Jitter]{\includegraphics[width=106pt]{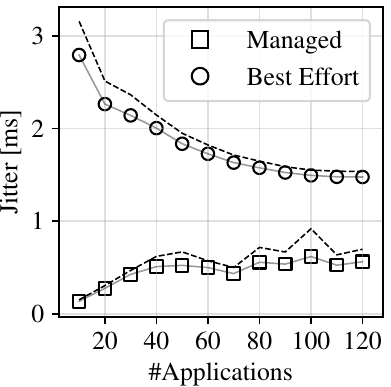}\label{fig:22_voip_metrics_jitter}}
%\subfigure[std(Delay)]{\includegraphics[width=105pt]{figures/24_delay_std.pdf}\label{fig:24_delay_std}}
\vspace{-10pt}
\caption{
Quality of Service metrics of the constrained link in terms of packet loss, delay and jitter for increasing number of applications ($|\Ax|$) as recorded by the VoIP clients.
The dashed lines without markers indicate the 95th percentile.
The markers indicate the managed (\managed{}) and best effort (\besteffort{}) median values.
Without data-rate management the queue at the bottleneck is overflowing quickly even at low numbers of parallel applications and thus causing packet loss and delay.
}
\label{fig:link_voip_qos}
\Description[Packet loss and RTT increases by number of applications for best effort, jitter decreases.]{No increase of packet loss and RTT for increased number of applications in the managed case, but jitter increases slightly. In the best effort case all three metrics show opposite trend compared to managed.}
\end{figure} 

Figure~\ref{fig:22_voip_metrics_loss} shows the median packet loss as measured by the VoIP clients during a call for 10 to 120 parallel applications.
The dashed lines indicate the 95th percentile.
The figure shows that there is no packet loss for the investigated number of applications in the managed experiments.
In the best effort experiments, the packet loss increases linearly from \perc{0.5} to \perc{7.1} (\perc{0.9} to \perc{8.1} for the 95th percentile).

Figure~\ref{fig:24_delay_delay} shows the Round-Trip-Time (RTT).
Note that the client-to-server flow direction of the constrained link is only lightly utilized and therefore, the given RTT approximates the one-way delay experienced by the applications.
In the best effort case, the delay increases roughly logarithmic from \ms{53} for 10 applications and saturates for 70 parallel applications at \ms{79}.
The 95th percentile shows that even with 10 parallel applications the experienced RTT is in \perc{5} of the cases already greater than \ms{79}.
In the managed experiments the measured RTT increases linearly from \ms{0.9} to \ms{1.1} (\ms{1.5} to \ms{2.4}).

Figure~\ref{fig:22_voip_metrics_jitter} shows the median and 95th percentile of the average jitter as measured by the VoIP clients during a call.
In general, the figure shows that in the best effort case the jitter decreases for increasing application count, while for the managed experiments the jitter increases.
The decrease in jitter in the best effort case shows that due to the link saturation, there are almost constant inter-arrival times of packets.
The high link utilization results in a full link queue and packets are processed at line-rate by the switch's outgoing interface.
In the managed case, the arrivals of the multiplexed requests of the clients result in minor RTT variations, but even for 120 applications the 95th percentile of the jitter stays below \ms{0.9}.

As there are no retransmissions for VoIP, the maximum delay for the successful transmission of a voice sample is about \unit[80]{ms} in our set-up.
For \perc{8} packet loss and \ms{80} delay, the utility for VoIP is estimated as 4.9 ($U_{VOIP}(80, 0.08) = 4.9$).
Hence, as defined by utility function $U_{VOIP}$, there is a maximum utility difference of 0.1 in the set-up (5 - 4.9).

In summary, data-rate management significantly improves the QoS metrics of the constrained link.
There is no packet loss, the RTT stays in most cases far below \ms{2.5} and the jitter is at least halved.
Regarding the influence of the QoS metrics on the VoIP utility, the VoIP clients in combination with the selected audio codec are marginally affected by the unmanaged link degradation.
However, one can imagine how applications with stricter QoS requirements or VoIP calls with longer network paths profit from the QoS improvements.

%Figures 
%\ref{fig:22_voip_metrics_loss} and \ref{fig:22_voip_metrics_loss} present the median packet loss and jitter over all requests of all VoIP applications.
%In the case of the managed experiments (squares in the figure), there is no packet loss in the experiment runs and the observed jitter is less than \unit[0.5]{ms}.
%For the best effort scenario (circles), packet loss and jitter increase with the number of simultaneous applications.
%For 10 parallel applications, the packet loss is less than \unit[1]{\%}.
%The loss increases roughly linear with the number of applications, up to \unit[6.3]{\%} for 120 applications.
%The jitter starts with about \fixme{x} at 10 parallel applications and \fixme{...}

%\subsection{Managed Gain for $|\Ax| = \{10, 20, .., 120\}$}\label{subsec:mgnt_gain}

\subsection{Increasing Number of Applications}\label{subsec:mgnt_gain}

\begin{figure}[t]
\centering
\subfigure[\vod{}]{\includegraphics[width=97pt]{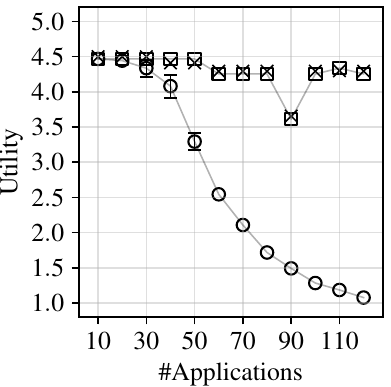}\label{fig:21_per_app_std_model_video}} %\hspace{15pt}
\subfigure[\live{}]{\includegraphics[width=97pt]{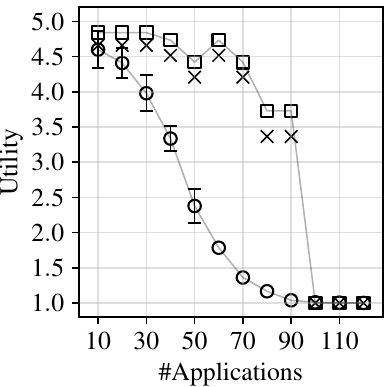}\label{fig:21_per_app_std_model_video_live}} %\hspace{15pt}
\subfigure[\ssh{}]{\includegraphics[width=97pt]{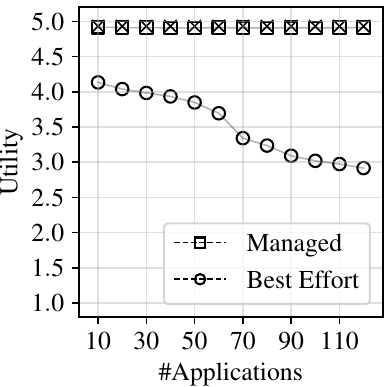}\label{fig:21_per_app_std_model_ssh}}
\subfigure[\web{}]{\includegraphics[width=97pt]{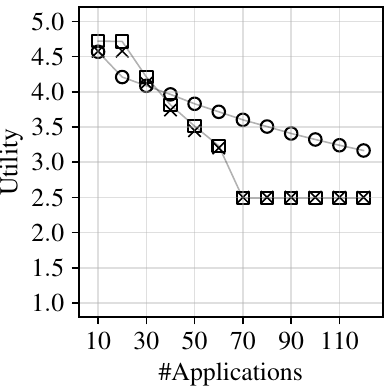}\label{fig:21_per_app_std_model_web}} \\ %\hspace{15pt}
\subfigure[\dl{}]{\includegraphics[width=105pt]{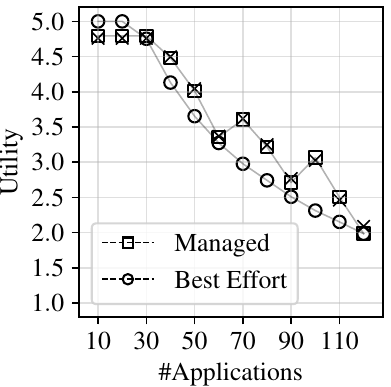}\label{fig:21_per_app_std_model_webdl}} \hspace{10pt}
\subfigure[Gain for all types]{\includegraphics[width=155pt]{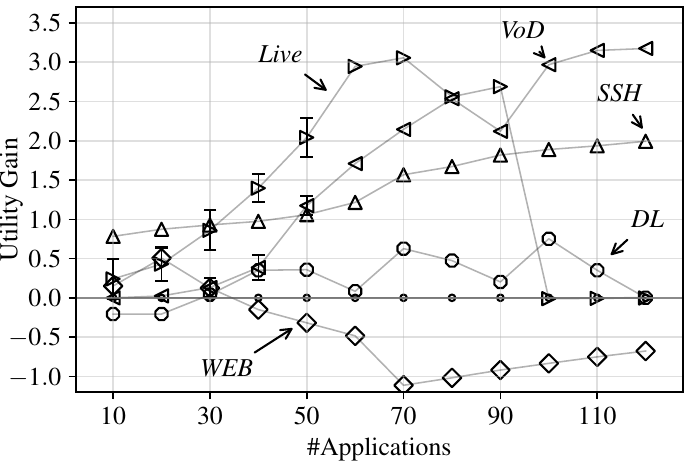}\label{fig:21_per_app_std_all_sum}} \hspace{10pt}
\subfigure[min]{\includegraphics[width=105pt]{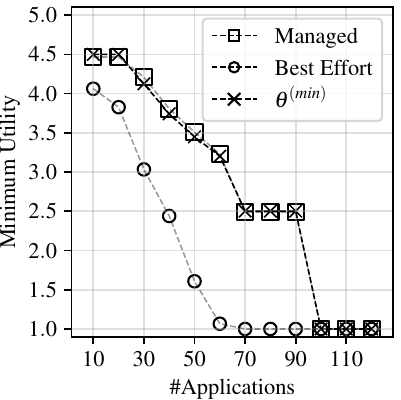}\label{fig:29b_min_utility}}
%\subfigure[VoIPClient]{\includegraphics[width=0.46\columnwidth]{figures/21_per_app_std_VoIPClient.pdf}\label{fig:02_dash_model}}
\vspace{-10pt}
\caption{
Comparison of managed measurements (\managed{}), target utility (\target{}) and best effort (\besteffort{}) measurements per application type for increasing number of applications sharing the constrained link.
In Figure~\subref{fig:29b_min_utility}, the crosses and the dashed line indicate the solution to the allocation formulation ($\UVMIN$) over all types.
Figure~\subref{fig:21_per_app_std_all_sum} summarizes differences in measured utility between best effort and managed.
Results are shown over mean of the \unit[10]{\%} tail of all requests of an application.
}
\label{fig:21_per_app_std}
\Description[Utility decreases faster for increased number of applications in the best effort case.]{Utility in the managed case stays longer above the best effort case for increasing numbers of applications. }
\vspace{-5pt}
\end{figure}

Figure~\ref{fig:21_per_app_std} illustrates the gain in utility per application type for increasing number of simultaneous applications.
%\rv{
Results are shown as the mean of the \unit[10th] percentiles of the utility values over all requests of an application. 
The \unit[10]{\%} tail as summary metric is chosen to allow for a small budget of random error compared to the minimal utility over all requests, e.g., for random delays in processing on the experiment PCs or requests which take longer due to rare latency spikes in the network.
%}
% ($mean( \{PERCENTILE_{10\%}(R_a) : \forall a \in A_t\} )$).
Hence, on average \unit[90]{\%} of the requests of a client result in a utility equal or better than the given value.
%The error bars indicate the standard deviation. %($std( \{PERCENTILE_{10\%}(R_a) : \forall a \in A_t\} ))$) in cases where it is at least 0.1.

Figures~\subref{fig:21_per_app_std_model_video} to \subref{fig:21_per_app_std_model_webdl} present the findings per application type.
The application class VoIP is omitted as there is no significant difference between the managed and best effort scenario.
Figure~\ref{fig:21_per_app_std_all_sum} summarizes the difference in utility per application type between the managed and best effort experiments.
Application types with a positive difference (top half of the figure) profit from management. 
The performance of application types with negative differences deteriorate.
The following general observations can be made based on the figures.

First, the utility for all shown types decreases with increasing number of applications in the best effort case.
This is expected as with increasing $|\Ax|$ more flows compete for the scarce constrained link capacity.
In the managed case, only \dl{} and \web{} exhibit an equivalent degradation in utility.
\vod{}, \live{}, and \ssh{} on the other hand can sustain a high utility in the managed experiments even while the number of competing flows increases.
Second, for $|\Ax| < 40$ the potential gain is low as the available capacity is sufficient to reach close to maximum utility for all applications in the managed and best effort cases.
Third, the performance of \web{} deteriorates while all other classes (except \voip{}) profit for most of the evaluated values of $|\Ax|$.
Fourth, the minimum utility over all applications ($\UVMIN$) in the managed case is mostly determined by \web{} and \live{}. %Starting from 110 applications \live{} decreases $UV_{min}$ sharply.
The minimum for the best effort case is mostly dictated by \ssh{} for $|\Ax| < 30$ and by \live{} for $|\Ax| \geq 30$.
Fifth, the measurements from the managed scenario deviate less than 0.5 from the target utility as determined by the solution to the optimization formulation for all application types.
For \dl{}, \web{}, \vod{}, and \ssh{} the deviation is even less than 0.2 for the investigated number of parallel applications.
Hence, data-rate management leads to predictability of application performance.
Furthermore, the results show that pacing can implement the output of the allocation optimization formulation accurately.

Next, we investigate the measurement results for each application type in detail.
For \textbf{\vod{}}~(Fig.~\ref{fig:21_per_app_std_model_video}), the utility decreases approximately linear with an increasing number of parallel applications for $|\Ax| > 30$. 
For $|\Ax| \leq 30$, best effort management is sufficient to provide a utility of 4.5 or higher.
With data-rate management, the fairness formulation can allocate enough resources to the \vod{} clients to sustain a high utility value even for up to $|\Ax| = 120$.
Hence, for $|\Ax| = 120$ the utility gain is about 3.1.
For \textbf{\live{}}~(Fig.~\ref{fig:21_per_app_std_model_video_live}), the figure shows that the utility decreases rapidly without data-rate management.
There, data-rate management is most effective at 60 to 70 parallel applications where the increase is up to 3.4.
In terms of predictable performance, the target utility is met most of the time with a deviation of 0.1 to 0.3.
However for $|\Ax| \geq 100$, the fairness formulation decreases the utility target to the minimum of 1.0, which is the same low utility as \live{} reaches in the best effort case for the same number of applications.
For \textbf{\ssh{}}~(Fig.~\ref{fig:21_per_app_std_model_ssh}), profit increases roughly linear with $|\Ax|$, from about 0.7 up to 2.1 for $|\Ax| = 120$.
Data-rate management avoids bursts and keeps the total data-rate under the constrained link capacity.
Hence, there is little queuing delay and the delay-sensitive applications like \ssh{} can sustain a high utility even for large $|\Ax|$.

For \textbf{\web{}}~(Fig.~\ref{fig:21_per_app_std_model_web}), the difference between managed and best effort is 1 or less utility (maximum difference of 0.9 at $|\Ax| = 90$). 
The target utility is close to the measured managed utility.
For $|\Ax| < 90$ and $|\Ax| > 90$ the difference decreases. 
As our pacing applies on application level, not flow level, \web{} can not gain an unfair advantage by opening multiple TCP connections anymore.
Furthermore, the utility function of \web{} (Fig.~\ref{fig:06_util_grid_web}) shows that \web{} is expensive in terms of required throughput, which makes the optimization likely to sacrifice the target utility of \web{} in the second optimization step in order to increase the average utility of all applications.
\textbf{\dl{}}~(Fig.~\ref{fig:21_per_app_std_model_webdl}) exhibits the smallest utility gains (besides \voip{}). 
The gain is below 0.8 for $|\Ax| \leq 90$ and around zero for $|\Ax| = 100$. 
The decrease of utility with increasing $|\Ax|$ is roughly linear for the managed and best effort experiments.
For $|\Ax| \geq 100$, the solution to the fairness problem increases the utility target for \dl{} again, which results in a utility gain close to 1.0.
Managing the utility is accurate and the deviation from the target utility can be neglected for all investigated numbers of parallel applications.
\textbf{\voip{}}~exhibits no benefit or degradation from the activated management according to the user experience model (further discussed in Section~\ref{subsec:voip}).

Figure~\ref{fig:29b_min_utility} shows the minimum \unit[10th] percentile utility as measured in the best effort and managed experiments and as calculated by the fairness formulation. % ($min( \{PERCENTILE_{10\%}(R_a) : \forall a \in A_t\} )$
The figure shows that in the managed scenario, every client's utility is at least 3.0 up to 80 parallel applications, which is denoted as \textit{fair} on the MOS scale.
In the best effort case, the observed minimum utility drops below 3 for 40 applications and down to 1.0 for 80.
When comparing $\UVMIN$ (\target{}) and managed (\managed{}), the managed minimum utility does not differ more than 0.1 from the calculated minimum utility.

In summary, the presented measurements for increasing number of parallel applications sharing the constrained link highlight the benefits of the proposed approach.
\vod{}, \live{}, \dl{}, and \ssh{} exhibit gains in utility between 0.5 and up to 3.3, even for 100 and more applications sharing the \mbps{100} link.
\web{}'s utility degrades, but the decrease is less than 1.0.
The minimum utility $\UVMIN$ can be greatly increased, especially for $|\Ax| > 30$, and the target utility is mostly met, resulting in predictable application performance.
\voip{} shows no benefit or degradation due to the nature of its user experience model.

\subsection{QoE Fairness}

To the best of our knowledge, there is no fairness measure to quantify the fairness for different application types with orthogonal resource demands, e.g., throughput-sensitive and delay-sensitive demands.
For example, \voip{} is in our set-up always close to a utility of 5.0, independent of other applications.
Hence, any fairness measure which considers only differences between values will consider this as unfair.
But enforcing equal utility for all application types, including artificially restricting \voip{}, would result in a non-Pareto-optimal utility distribution where the target utility of \voip{} could be increased without negatively impacting other applications.
Therefore, we evaluate the inter-application fairness per application type.
Note that for the evaluation we are restricting the allocation formulation to allocate only one target utility value per application type.
Hence the target utilities per type exhibit always perfect fairness and are omitted.

\begin{figure}[t]
\centering
\subfigure[\vod{}]{\includegraphics[width=73pt]{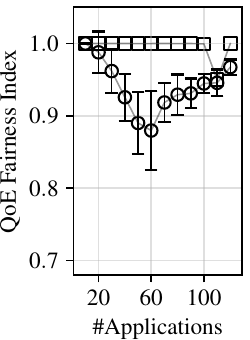}\label{fig:29_fairness_model_video}} \hspace{4pt}
\subfigure[\live{}]{\includegraphics[width=73pt]{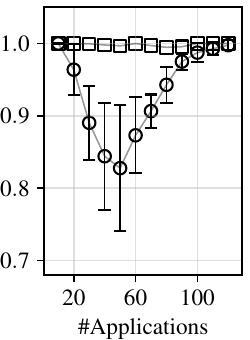}\label{fig:29_fairness_model_video_live}} \hspace{4pt}
\subfigure[\ssh{}]{\includegraphics[width=73pt]{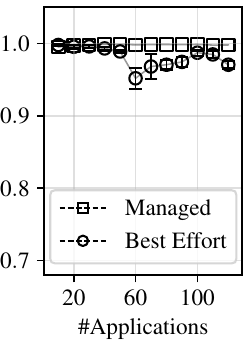}\label{fig:29_fairness_model_ssh}} \hspace{4pt}
\subfigure[\web{}]{\includegraphics[width=73pt]{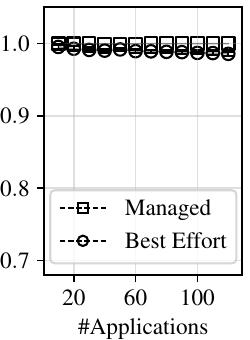}\label{fig:29_fairness_model_web}} \hspace{4pt}
\subfigure[\dl{}]{\includegraphics[width=73pt]{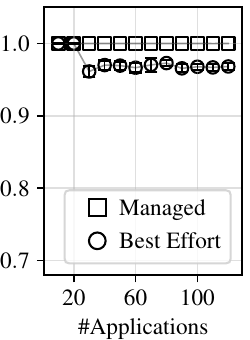}\label{fig:29_fairness_model_webdl}}
%\subfigure[VoIPClient]{\includegraphics[width=0.46\columnwidth]{figures/21_per_app_std_VoIPClient.pdf}\label{fig:02_dash_model}}
\vspace{-1em}
\caption{
Comparison of the F-index between managed (\managed{}) and best effort (\besteffort{}) scenarios per application type for increasing number of applications sharing the constrained link.
An F-index of 1.0 denotes perfect fairness.
}
\label{fig:29_fairness}
\Description[Fairness of best effort below fairness of managed.]{High fairness for managed case. First decreasing and then increasing fairness for best effort case for increasing number of applications.}
\vspace{-1em}
\end{figure}

We evaluate the inter-application fairness using the F-index \cite{hossfeld2017definition} defined by $F=1 - \frac{2\sigma}{4}$ for a utility scale of 1 to 5.
The F-index is selected as fairness measure as it is specifically designed and evaluated for user experience fairness.
An F-index of 1.0 indicates perfect fairness between the applications.
An F-index of 0.0 is the result of half of the application experiencing a utility of 1.0 and the other half a utility of 5.0.
Figures \ref{fig:29_fairness_model_video} to \ref{fig:29_fairness_model_webdl} illustrate the F-index per application type $t$ for $|\Ax| = \{10, 20, .., 120\}$ applications sharing the constrained link, measured for the best effort and managed scenarios. %($F(\{PERCENTILE_{10\%}(R_a) : \forall a \in A_t\})$).
From the figures, we conclude that in the managed case, the F-index does not drop below 0.98 for any of the evaluated scenarios and application types.

In the best effort case, the fairness depends strongly on the application type and the number of parallel applications.
The \web{} clients exhibit a fairness similar to that in the managed scenario ($\geq 0.98$).
For \ssh{} and \dl{}, the fairness fluctuations are larger, but in general the fairness is still high ($\geq 0.95$).
The two video streaming types \vod{} and \live{} suffer the most in the best effort scenarios.
For \live{}, the fairness drops down to 0.7 for $|\Ax|=44$ and for \vod{} down to 0.77 for $|\Ax|=55$.
However, for video streaming there is a high level of fairness for $|\Ax| < 30 $ and $|\Ax| > 100$.
This is due to the fact that for low number of parallel applications, there is sufficient capacity for all clients to reach close to maximum utility while for a high number of parallel applications all clients are close to a utility value of 1.0. 

In summary, the evaluation of the fairness per application type shows that \vod{} and \live{} profit the most from the management.
\ssh{} and \dl{} show some improvement. \web{} and \voip{} improve only marginally.
In the managed measurements, we observe nearly perfect fairness for all application types.

\subsection{Summary}

The evaluation set out to discuss the following four subjects:
\begin{enumerate*}[label=\roman*)]
\item comparison of minimum and average utility for managed and best effort scenarios,
\item advantages and disadvantages of central data-rate management for each application class,
\item predictability of application performance, and
\item fairness between the applications.
\end{enumerate*}

First, a scenario with 80 applications sharing a \unit[100]{Mbps} link is presented.
The measurements show that for the best effort case, web browsing consumes about four times more of the available throughput than the other applications.
This is due to web browsers using multiple parallel TCP connections.
As a consequence, the utility of the web browsing sessions is high (3.5 to 4.0), while other applications like live video streaming suffer ($\leq 2$).
Next, the allocation formulation is solved for the 80 applications and pacing is applied to the applications.
The results show that video streaming, remote terminal work, and file download can increase their utility by 1 to 3 while web's utility is only decreased by 1.
Furthermore, the standard deviation of a client's utility is decreased to $\leq 0.1$ from 0.2 to 0.8 in the best effort case, resulting in predictable application performance.
The measurements for 10 to 120 parallel applications sharing the link support the findings of the 80 applications scenario.
%In general video streaming, ssh and web download benefit from the management by more than 1 utility while web browsing's utility is decreased by less than 1.
The evaluation of the fairness shows that in the managed scenarios, the application types exhibit close to perfect fairness.
For the best effort scenarios, the fairness results show that the two video streaming intents profit the most from the management, followed by \dl{} and \ssh{}.
The \web{} clients do not profit much from the management in terms of fairness.

No or little benefit can be expected from the management when the link is only lightly utilized, as the applications do not have to compete for resources and there is no queuing time at the bottleneck.
For highly utilized links, throughput-sensitive applications can not profit as the available resources are insufficient for all applications.
In such situations some application could be evicted from the network to provide a satisfying experience for critical applications.
However, delay-sensitive applications like \ssh{} still profit from the reduced queuing at the link.

In summary, the results show that there is a significant benefit of centrally controlled application pacing in terms of utility, inter-application fairness, and predictability.
Furthermore, compared to classical Quality of Service measures in the network, the approach can be implemented with heterogeneous forwarding devices without any special features, it does not require expensive switch buffer space, and it is fully software-based.

%% file: content/10_conclusion.tex
\section{Conclusion} \label{sec:conclusion}  

%Achieving scalable and application-aware multi-application resource allocation in enterprise networks is challenging.
%By default applications can send packets into the network at will.
%Configuring traffic policing or shaping on forwarding devices in the network leads to queuing and packet loss, which introduces delay and decreases transmission efficiency.

In this paper we propose a design for resource allocation in enterprise networks based on central software-defined network control, fine-grained per-application pacing at the end-hosts, and utility functions derived from measurements and user-experience models.
Pacing refers to the method of restricting the amount of data an application is allowed to send into the network by implementing local back-pressure to the application sockets and introducing artificial delays between packets.
Traditional methods of QoS control in the network, such as policing or scheduling, interact badly with end-host congestion control and do not scale to larger number of applications and application classes.
Moving application pacing from in-network QoS methods to the end hosts, e.g., to user PCs, servers, smartphones, and tablets, is scalable, increases transmission efficiency, reduces the required complexity of forwarding devices, and allows cost-efficient high link utilizations.
To the best of our knowledge, this is the first work proposing, formulating, and evaluating a scalable architecture for resource allocation for end-user applications in enterprise environments based on real applications and user-experience models.

We define application- and user-level utility using selected user-experience models from the literature.
Based on the models, we derive per-application utility models for the five common network use cases web browsing, file download, remote terminal, adaptive video streaming, and Voice-over-IP.
Afterwards, we determine sensible resource allocations by formulating a two-stage mixed-integer linear program based on the number and types of applications, their utility functions, and network resources.
The mixed-integer linear program decides on how to embed the applications in the network in terms of the allowed data-rate per application and the delay-constrained routing of the application flows.
Once the allowed rate and routing is determined, the flow routing is configured through an SDN protocol and the pacing is enforced through local agents at the end-hosts.

We evaluate the methodology by implementing a proof-of-concept testbed with a throughput-constrained link and an increasing number of parallel applications sharing the link.
The results show that QoS metrics, such as delay and packet loss, considerably improve with pacing, due to the controlled link utilization.
%Furthermore, when looking at the fairness in terms of utility, the results show that the minimum and total utility over all applications is significantly improved.
When looking at the fairness per application type, the results show that there is near perfect fairness between the clients.
For the five evaluated application types, the results show that web browsing's utility decreases, as it has an unfair advantage in the best effort case due to its multiple parallel TCP-connections.	
However, the loss in utility of web browsing is low, compared to the gain for the other types.
Real-time applications, such as remote terminal work, profit due to the reduced delay and packet loss.
%Adaptive video streaming and file download benefit from the stable and predictable goodput.
VoIP enjoys lower packet loss, delay, and jitter, but does not suffer in terms of utility by one impaired link due to the resilient audio codec.
From the experiments, we conclude that the proposed architecture enables scalable resource allocation and predictable application performance.

This paper is a step towards extending Software-defined Networking towards the edge of the network with scalable resource allocations from the perspective of the human users.
Future work in this area should focus on how to autonomously create and update utility functions, investigate the impact of inaccurate utility functions, develop fast heuristics for the allocation problem formulation, evaluate further application types, and solve the problem of dynamically recomputing pacing rates and embedding additional applications at run-time.

\begin{acks}
This work has been supported, in part, by the German Research Foundation (DFG) under the grant numbers KE1863/6-1, ZI1334/2-1 and TR257/43-1
and in part by the European Union's Horizon 2020 research and innovation program (grant agreement No 647158 - FlexNets).
This work reflects only the authors' view and the funding agency is not responsible for any use that may be made of the information it contains.
\end{acks}

%% file: content/11_appendix.tex
\appendix

\section{Allocation Problem Formulation} \label{appendix:problem}

Next we give the complete description of the resource allocation problem formulated as a MILP.
The MILP has to consider the two-dimensional utility function of every application, the capacities of all links and the delay on intermediate links depending on the link utilization.
The decision variables describe which pacing rate to apply to which application and how to configure the routing between application endpoints.
The problem can be summarized with the following inputs, objectives, high-level constraints and outputs.

\begin{enumerate}[leftmargin=51pt,labelindent=16pt]
\item[\textbf{Inputs:}] (I) Number of applications. (II) Utility function $\Ux$ of each application. (III) Network topology with link capacity information and delay on the links based on link utilization. 
\item[\textbf{Objectives:}] (I) Min-max utility fairness in the first step. (II) Increasing average utility in the second step.
\item[\textbf{Constraints:}] Unidirectional application routing (source to destination) has to be valid, considering link capacity and maximum delay per application.
\item[\textbf{Outputs:}] (I) Target utility value and allocated throughput per application. (II) Application flow routing.
\end{enumerate}  

\subsection{Notation}

Table~\ref{tab:notation} summarizes the notation. 
$\Ax, a \in \Ax$ is the set of all unidirectional application flows $a$.
For simplification, application flow $a$ and intent $i$ are merged in the notation to only $a$ and each application consists of only one application flow.
The two directions of a bidirectional application flow are considered as two independent applications by the formulation.
This allows different paths and utility functions for both flow directions.
We define the topology as a directed graph $G(\mathcal{V},\mathcal{E})$ with nodes $v \in \mathcal{V}$ and edges $(u, v) \in \mathcal{E}$ and edge capacity $\Cx_{u, v}$.
A flow $a$ is defined by the source node $\Sx_a$, target node $\Tx_a$ and its utility function $\Ux$.

$\DCU$ and $\DCD$ (both $\in \mathbb{R_+}^{|\mathcal{E}| \times |\mathcal{E}| \times m}$) describe the piece-wise defined relationship between link usage $\DCU$ and delay $\DCD$ for specific edge and for a quantization bin $m$.

The utility function describes the relationship between allocated throughput and delay and the application's resulting utility (Fig.~\ref{fig:util_grids}).
It can be determined for example through measurements and user experience models, as we do in the paper at hand in Chapter~\ref{sec:applications}.
Mathematically, the utility function is split into its three components, the throughput demands ($\UTP$ $\in \mathbb{R_+}^{|\Ax| \times n}$), the delay demands ($\UD$ $\in \mathbb{R_+}^{|\Ax| \times n}$) and the utility values ($\Ux$ $\in ([1, 5])^{|\Ax| \times |\UTP| \times |\UD|}$), where $n$ denotes the quantization bin. 

$\Fx$ describes the application flow routing. 
An edge $(u, v)$ is traversed by an application $a$ if $\Fx_{a, u, v}$ equals $1$.
Delay on a link is describes as a function of the link usage.
$\DCU$ and $\DCD$ (both $\in \mathbb{R_+}^{|\mathcal{E}| \times |\mathcal{E}| \times m}$) define the piece-wise defined relationship between usage ($\DCU$) and resulting delay ($\DCD$) for each edge in the graph $G$ and quantization bin $m$.

\begin{table}%[]
\vspace{2em}
	\caption{Notation Allocation Problem Formulation}
    \begin{tabular}{llll}
%    \hline
    Symbol & Type & Unit & Description \\ \hline %\hline

    \multicolumn{4}{|c|}{Constants} \\ \hline
    $G(\mathcal{V},\mathcal{E})$ & & & Network topology graph with nodes $\mathcal{V}$ and edges $(u,v) \in \mathcal{E}$. \\ 
    $\Ax, a \in \Ax$ & & & Set of all unidirectional application flows. \\
    $\Sx$, $\Tx$  & $\in \mathcal{V}^{|\Ax|}$ & & Start and target nodes of application flows. \\
    $\DCU$, $\DCD$ & $\in \mathbb{R_+}^{|\mathcal{E}| \times |\mathcal{E}| \times m}$ & & Translation between link usage and delay for a specific link. \\
    $\Cx$ & $\in \mathbb{R_+}^{|\mathcal{V}| \times |\mathcal{E}|}$ & Kbps & Unidirectional link capacity between $u$ and $v$. \\ 
    $\UTP$ & $\in \mathbb{R_+}^{|\Ax| \times n}$ & Kbps & Utility functions' throughput demands of the applications. \\
    $\UD$ & $\in \mathbb{R_+}^{|\Ax| \times n}$ & ms & Utility functions' delay demands of the applications. \\ 
    $\Ux$  & $\in ([1, 5])^{|\Ax| \times |\UTP| \times |\UD|}$ & & Utility functions' utility values of the applications. \\
%    $\tp$ & $\in \{0, 1, .., |\BTP|\}^{|\mathcal{A}|}$ & Index into $\BTP_a$ and $\UTP_a$. \\
%    $\dx$ & $\in \{0, 1, .., |\BD|\}^{|\mathcal{A}|}$ & Index into $\BD_a$ and $\UD_a$. \\ \hline
    \hline
    \multicolumn{4}{|c|}{Decision Variables} \\ \hline    
    $\UVMIN$ & $\in [1, 5]$ & & Minimum utility for all applications. \\
    $\BTP$ & $\in \{0, 1\}^{|\mathcal{A}| \times |\UTP|}$ & & 1 if a specific throughput demand index is selected for an application. \\ 
    $\BD$ & $\in \{0, 1\}^{|\Ax| \times |\UD|}$ & & 1 if a delay demand index for application is selected. \\ 
    $\Fx$ & $\in \{0, 1\}^{|\Ax| \times |\mathcal{E}| \times |\mathcal{E}|}$ & & 1 if an edge is traversed by an application. \\ \hline    
    
    \multicolumn{4}{|c|}{Functions} \\ \hline    
    
%    $tp^a, d^a$ & Throughput and delay index chosen for application $a$ \\
    $\TP(a)$ & $\Ax \mapsto \mathbb{R_+}$ & Kbps & Selected throughput for application $a$. \\
    $\Dx(a)$ & $\Ax \mapsto \mathbb{R_+}$ & ms & Selected delay requirement for application $a$. \\
    $\TU(a)$  & $\Ax \mapsto [1, 5]$ & & Target utility value of application $a$. \\
    $\LU(u,v)$ & $\mathcal{E} \mapsto \mathbb{R_+}$ & Kbps & Assigned throughput to link $(u, v)$ in Kbps. \\
    $\LD(u,v)$ & $\mathcal{E} \mapsto \mathbb{R_+}$ & ms & Delay on link $(u, v)$ in milliseconds. \\
    $\AD(a)$ & $\Ax \mapsto \mathbb{R_+}$ & ms & End-to-end delay of application $a$ in milliseconds. \\ \hline
    %$B\_U^a_{tp_a,d_a}$ & Binary matrix to select utility value from $U^a$ \\ \hline    

    \multicolumn{4}{|c|}{Miscellaneous} \\ \hline
    $\UV^{(\text{min,\{1|2\}})}$ & $\in [1, 5]$ & & Solution of $\UV^{\text{(min)}}$ in first and second step. \\
    $\epsilon$ [$=0.3$] & $\in \mathbb{R^+}$ & & Slack parameter for $\UV^{\text{(min)}}$ in the second step. \\ 
    $n$, $m$ & $\in \mathbb{N}$ & & Quantification factors for the utility and link delay functions. \\ 
    \end{tabular}
\label{tab:notation}
\end{table}

\subsection{Objective}

The objective of the MILP is in the first step to maximize the minimal utility value $\UVMIN$ over all applications.
In the second step the MILP maximizes the sum of all utility values while the minimum utility is $\UVMIN$ restricted to range based on the minimum value determined by the first step, denoted as $\UVMINI$, $\UVMIN \in [\UVMINI - \epsilon, \UVMINI]$ with $\epsilon = 0.3$.
The second step allows the problem formulation to improve the average utility over all applications by relaxing the max-min fairness constrain using the slack parameter $\epsilon$.
This prevents solutions where the optimization would stop when the utility of a single application can not be increased further, but where there are plenty of resources left to increase the utility of other applications.

We define $\UV_a$ as utility value of an application $a$.
In the first step we maximize the minimum utility value (\textit{max-min fairness}) subject to all application utilities have to be larger than the minimum utility value $\UVMIN$:

\vspace{-0.5em}
%\begin{equation}
\begin{align}
\text{maximize:} & \quad \UVMIN \\
\label{eq:obj_step1}
\text{subject to:} & \quad \TU(a) \geq \UVMIN  \quad \forall a \in \Ax \\
                   & \quad \text{and (\ref{eq:a_tp}) - (\ref{eq:last_delay_constr})}
\end{align}
%\end{equation}

\noindent
We denote the optimal value of $\UVMIN$ of the first step as $\UVMINI$.
In the second step we relax the max-min constraint by $\epsilon$ and maximize the sum of all utility values.
We denote the optimal value of $\UVMIN$ of the second step as $\UVMINII$ and add the additional constraint to bound $\UVMINII$ by $\UVMINI - \epsilon = 0.3$:

%\vspace{-0.9em}
%\begin{equation}
%\text{maximize:} \quad \sum_{a \in \Ax} U(a)
%\end{equation}

\vspace{-0.5em}
%\begin{equation}
\begin{align}
\text{maximize:} & \quad \sum_{a \in \Ax} \TU(a) \\
\label{eq:obj_step2}
\text{subject to:} & \quad \UVMIN \geq \UVMINI - \epsilon \\
                   & \quad \text{and (\ref{eq:a_tp}) - (\ref{eq:last_delay_constr})}
\end{align}
%\end{equation}
\vspace{0.05em}

For remainder of this formulation and if not otherwise stated, $\UVMIN$ denotes the optimal value as determined by the second step ($\UVMINII$).
Next we formulate the constraints.
Table~\ref{tab:contraints} summarizes the constraints.

\begin{table}[b]
\caption{Overview of all constraints}
\begin{tabular}{|l|l|p{9cm}|}
\hline
Type     & Constraints & Description                                                                                                 \\ \hline
Objectives & (\ref{eq:obj_step1}), (\ref{eq:obj_step2})  & Maximize minimum utility (1st step) and sum of utilities (2nd step). \\
Utility  & (\ref{eq:a_tp}) - (\ref{eq:u_a})  & Select target utility, throughput allocation and maximum allowed delay per application.                     \\
Routing  & (\ref{eq:fc}) - (\ref{eq:fc_t}) & Application routing (multi-commodity flow problem).  \\
Capacity & (\ref{eq:lusage}) - (\ref{eq:capacity_constrain}) & Link capacity (in Kpbs) can not be exceeded by applications. \\
Delay    & (\ref{eq:Sp}) - (\ref{eq:last_delay_constr}) & Determine delay per link (in milliseconds) depending on link usage. Ensure applications' maximum delay demand is not exceeded. \\ \hline
\end{tabular}
\label{tab:contraints}
\end{table}

\subsection{Utility Selection Constraints}\label{subsec:appendix_utilselection}

%First we define the selection of throughput, delay and utility value for each application.
For each application, one throughput, delay and target utility value have to be selected.
We first introduce the equations and afterwards illustrate the selection process by a simplified example.
Eq.~\ref{eq:a_tp} and Eq.~\ref{eq:a_d} dictate that only one throughput and delay demand for application $a$ can be chosen at a time: 

\begin{equation}
\label{eq:a_tp}
\sum^{|\BTP_a|}_{i=1} \BTP_{a,i} = 1 \quad \forall a \in \Ax
\end{equation}
\begin{equation}
\label{eq:a_d}
\sum^{|\BD_a|}_{i=1} \BD_{a,i} = 1 \quad \forall a \in \Ax
\end{equation}

\vspace{1em}

\noindent
Hence the chosen throughput demand in Kbps $\TP^a$ and delay requirement in milliseconds $\Dx^a$ for application $a$ are given by the following element-wise multiplications.

\begin{equation}
\TP(a) := \BTP_a^T \cdot \UTP_a
\end{equation}

\begin{equation}
\Dx(a) := \BD_a^T \cdot \UD_a
\end{equation}

\vspace{1em}

%We define $B\_U^a_{tp,d}$ as the utility decision as given by $B\_TP^a$ and $B\_D^a$ for an application $a$.
%The index of the utility value is set to one through the following constrains:
%\begin{equation}
%\forall a \in A, \forall d : \sum_{tp=1}^{|B\_TP^a|} B\_U^a_{tp,d} = B\_TP^a_{tp}
%\end{equation}
%\begin{equation}
%\forall a \in A, \forall tp : \sum_{d=1}^{|B\_D^a|} B\_U^a_{tp,d} = B\_D^a_{d}
%\end{equation}

\noindent
The resulting utility value of application $a$, $\TU(a)$, is then selected from the quantified  utility functions (Fig.~\ref{fig:util_grids}) by the following equation:

\begin{equation}
\label{eq:u_a}
\TU(a) := \sum_{tp=1}^{|\BTP|} \sum_{d=1}^{|\BD|} (\BTP_{a,tp} \cdot \BD_{a,d} \cdot \Ux_{a,tp,d})
\end{equation}

\noindent
Next we give an example for a target utility, throughput and delay demands calculations for an arbitrary application $a$.
The discretized utility function $\Ux_a$ has a domain of [100, 500, 1000] Kbps for the throughput and [150, 100, 50] milliseconds for the delay demand.
At an allocation of \unit[1000]{Kbps} and \unit[50]{ms} the utility of the application reaches its highest point with 4.9, while for \unit[100]{Kbps} and \unit[150]{ms} the target utility drops to $1.3$.
In the following example the decision variables $\BTP_{a,1}$ and $\BD_{a,1}$ are set to $1$ by the solver based on other constraints like the available link capacity.
Hence, an allocation of $\TP(a) = \unit[500]{Kbit/s}$ is chosen with a target utility of $\TU(a) = 3.0$.

\[
\TU(a) ~ = ~ \bordermatrix{~ & \BTP_{a,0}  & \BTP_{a,1}  & \BTP_{a,2} \cr
             \BD_{a,0} & \Ux_{a,0,0} & \Ux_{a,1,0} & \Ux_{a,2,0} \cr
             \BD_{a,1} & \Ux_{a,0,1} & \Ux_{a,1,1} & \Ux_{a,2,1} \cr
             \BD_{a,2} & \Ux_{a,0,2} & \Ux_{a,1,2} & \Ux_{a,2,2} \cr} ~
     = ~ ~ \bordermatrix{~ & 0 & \mathbf{1} & 0 \cr
                         0 & 1.3            & 1.6          & 2.1 \cr
                \mathbf{1} & 2.9            & \mathbf{3.0} & 3.5 \cr
                          0 & 4.2            & 4.3         & 4.9 \cr}
     = 3.0
\]

\[
\TP(a) ~ = ~ \begin{pmatrix}
			  \UTP_{a,0}  \\
			  \UTP_{a,1}  \\
			  \UTP_{a,2}  \\
 			\end{pmatrix}
 			\cdot
 			\begin{pmatrix}
			  \BTP_{a,0} & \BTP_{a,1} & \BTP_{a,2} \\
 			\end{pmatrix}
	~ = ~ \begin{pmatrix}
			   100  \\
			   \mathbf{500}  \\
			  1000  \\
 			\end{pmatrix}
 			\cdot
 			\begin{pmatrix}
			  0 & \mathbf{1} & 0 \\
 			\end{pmatrix}
 	~ = ~ \unit[500]{Kbit/s}
\]

\[
\Dx(a) ~ = ~ \begin{pmatrix}
			  \UD_{a,0}  \\
			  \UD_{a,1}  \\
			  \UD_{a,2}  \\
 			\end{pmatrix}
 			\cdot
 			\begin{pmatrix}
			  \BD_{a,0} & \BD_{a,1} & \BD_{a,2} \\
 			\end{pmatrix}
	~ = ~ \begin{pmatrix}
			   150  \\
			   \mathbf{100}  \\
			  50  \\
 			\end{pmatrix}
 			\cdot
 			\begin{pmatrix}
			  0 & \mathbf{1} & 0 \\
 			\end{pmatrix}
 	~ = ~ \unit[100]{ms}
\]

%\reffig{fig:vars_utility} illustrates the utility related variables by example.
%In the example, $B\_TP^a$ and $U\_TP^a$ define the selected throughput $TP^a = \kbps{250}$.
%$B\_D^a$ and $U\_D^a$ define the selected delay $D\^a = \unit[60]{ms}$.
%$B\_TP^a$ and $B\_D^a$ select an utility value of $UV^a = 4$.
%
%\begin{figure}[]
%\centering
%\includegraphics[width=160pt]{figures/utilgrid.pdf}
%\caption{
%Illustration of the utility related variables and constants. 
%$B\_D$ and $U\_D$ describe the chosen maximum allowed delay, $B\_TP$ and $U\_TP$ the throughput requirement for an application $a$. 
%$U^a$ with $B\_D^a$ and $B\_TP^a$ describes the resulting utility value for that application.
%}
%\label{fig:vars_utility}
%\end{figure}

\subsection{Routing Constraints}

We formulate the application flow routing problem as the multi-commodity flow problem \cite{hu1963multi} with non-fractional flows. 
First we formulate the constraints required to route the flow from source to destination.
Afterwards we formulate the link capacity and application delay constraints.
A flow is subject to the following routing constraints.
Number of incoming and outgoing edges of in-between nodes has to be equal (\textit{flow conservation}): 

\begin{equation}
\label{eq:fc}
\sum_{w \in \mathcal{V}} \Fx_{a,u,w} = \sum_{w \in \mathcal{V}} \Fx_{a,w,u}~~~~~~|~u \neq \Tx_a, \Sx_a \quad \quad \forall a \in \Ax, \forall u \in \mathcal{V}
\end{equation}
\vspace{0.5em}

\noindent
Flow conservation at the source (Eq.~\ref{eq:fc_s}) and destination (Eq.~\ref{eq:fc_t}):

\begin{equation}
\label{eq:fc_s}
\sum_{w \in \mathcal{V}} \Fx_{a,\Sx_a,w} - \sum_{w \in \mathcal{V}} \Fx_{a,w,\Sx_a} = 1 \quad \forall a \in \Ax
\end{equation}

\begin{equation}
\label{eq:fc_t}
\sum_{w \in \mathcal{V}} \Fx_{a, w, \Tx_a} - \sum_{w \in \mathcal{V}} \Fx_{a, \Tx_a, w} = 1 \quad \forall a \in \Ax
\end{equation}

\subsection{Capacity Constraints}

Capacity constraints ensure that the assigned throughput to a link does not exceed the capacity of the link.
Next we formulate the required link capacity constraints.
We define the link usage in Kbps $\LU(u,v)$ on the directed edge $(u,v)$ as the sum of the throughput values of all applications traversing that edge/link:

\begin{equation}
\label{eq:lusage}
\LU(u, v) := \sum^{}_{a \in \Ax} \Fx_{a,u,v} \cdot \TP(a)
\end{equation}

\noindent
And assigned throughput can not exceed the capacity:

\begin{equation}
\label{eq:capacity_constrain}
\LU(u,v) \leq \Cx_{u,v} \quad \forall(u,v) \in \mathcal{E}
\end{equation}

\subsection{Delay Constraints}

We define the delay of each link as a function of the link usage.
That way, the delay function can express a combination of constant, e.g., propagation delay, and dynamic, e.g., queuing and processing delay, use cases.
For example, an added constant delay can describe significant propagation delay, or the queuing delay can be modeled based on the target link utilization.
We first provide the necessary equations and then provide a simple example.

%\begin{figure}[h]
%\centering
%\includegraphics[width=0.97\columnwidth]{figures/delayfunc.pdf}
%\caption{Example approximated link delay calculation and variables}
%\label{fig:vars_delay}
%\end{figure}

We do a piece-wise linear interpolation to approximate the link delay for edge $(u,v)$, denoted as $\LD(u,v)$, for a given link usage $\LU(u,v)$ of the edge.
$\DCU_{u,v,i}$ and $\DCD_{u,v,j}$ describe the piece-wise defined translation sets between a usage in Kbps with index $i$ and delay in milliseconds with index $j$ for a link $(u,v)$ with $|\DCU_{u,v}|=|\DCD_{u,v}|$.
We introduce the variables $l_{u,v,p}$ with $l_{u,v,p} \in \{0, 1\}$ and $S^{u,v,p} \in [0, 1]$ for $p = \{0, 1, .., |\DCU_{u,v}|-1\}$.
Variable $l$ selects the closest, lower, link usage from $\DCU$ and $S$ is the linear scaling factor.
$l$ and $S$ are subject to:

\begin{equation}
\label{eq:Sp}
S_{u,v,p} \leq l_{u,v,p} \quad \forall (u,v) \in \mathcal{E},~~p = \{0, 1, .., |\DCU_{u,v}|-1\}
\end{equation}
\vspace{0.1em}

\noindent
Constrain the selection variable $l_{u,v,p}$ and scale variable $S_{u,v,p}$ according to the link usage $\LU_{u,v}$:

\begin{equation}\label{eq:constr_eq_lusage}
\LU(u,v) - \sum^{|\DCU_{u,v}|-1}_{p=0} [l_{u,v,p} \cdot \DCU_{u,v,p} + (\DCU_{u,v,p+1} - \DCU_{u,v,p}) \cdot S_{u,v,p}] = 0 \quad \forall (u,v) \in \mathcal{E}
\end{equation}
\vspace{0.1em}

\noindent
$\LD(u,v)$ then defines the delay for the given link usage:

\begin{equation}\label{eq:constr_eq_ldelay}
\LD(u,v) := \sum^{|\DCU_{u,v}|-1}_{p=0} [l_{u,v,p} \cdot \DCD_{u,v,p} + (\DCD_{u,v,p+1} - \DCD_{u,v,p}) \cdot S_{u,v,p}]
\end{equation}
\vspace{0.1em}

\noindent
Let's consider the following simple example.
A hypothetical link $(u,v)$ has a maximum capacity of \unit[1000]{Kbps} and a propagation delay of \unit[10]{ms}.
Up to a link usage of \unit[100]{Kbps}, there is no queuing delay.
Between \unit[100]{Kbps} and \unit[1000]{Kbps} the queuing delay increases linearly up to a maximum of \unit[70]{ms}. 
Hence, at a link usage of \unit[1000]{Kbps} the delay on the link is $\unit[70]{ms} + \unit[10]{ms} = \unit[80]{ms}$.
We can model this by setting $\DCU$ and $\DCD$ as follows:

%\begin{tikzpicture}[scale=5.0]
%
%% horizontal axis
%\draw[->] (0,0) -- (1.5,0) node[anchor=north] {Link Usage};
%% labels
%\draw	(0,0) node[anchor=north] {0 Kbps}
%		(0.1,0) node[anchor=north] {100 Kbps}
%		(1.0,0) node[anchor=north] {1000 Kbps};
%
%% vertical axis
%\draw[->] (0,0) -- (0,1) node[anchor=east] {Delay};
%% nominal speed
%\draw[dotted] (0.5,0) -- (0.5,1);
%
%% Us
%\draw[thick] (0,0) -- (0.5,0.2) -- (0.5,0.2) parabola[bend at end] (1.0,0.8);
%%\draw (1,1.5) node {$U_s$}; %label
%
%% Psis
%%\draw[thick,dashed] (0,3) -- (2,3) parabola[bend at end] (6,1);
%%\draw (2.5,3) node {$\varPsi_s$}; %label
%
%\end{tikzpicture}

\[
\DCU_{u,v} ~ = ~ \begin{pmatrix}
		 	  \DCU_{u,v,0}    \\
			  \DCU_{u,v,1}  \\
			  \DCU_{u,v,2} \\
 			\end{pmatrix}
		~ = ~ \begin{pmatrix}
		 	  0    \\
			  100  \\
			  1000 \\
 			\end{pmatrix} Kbps
 			\quad
\DCD_{u,v} ~ = ~ \begin{pmatrix}
		 	  \DCD_{u,v,0}    \\
			  \DCD_{u,v,1}  \\
			  \DCD_{u,v,2} \\
 			\end{pmatrix}
		~ = ~ \begin{pmatrix}
		 	  10 \\
			  10 \\
			  80 \\
 			\end{pmatrix} ms
\]

\noindent
Let us assume the decision variables assign link $(u,v)$ a total link usage of \unit[500]{Kbps}.
The resulting total delay on that link can then be calculated by first determining $l$ and $S$:
%For a specific link $(u,v)$:

\begin{align*}
\LU(u,v) &- \sum^{|\DCU_{u,v}|-1}_{p=0} [l_{u,v,p} \cdot \DCU_{u,v,p} + (\DCU_{u,v,p+1} - \DCU_{u,v,p}) \cdot S_{u,v,p}] = 0 \\
\leftrightarrow 500 &- ([l_{u,v,0} \cdot 0 + (100 - 0) \cdot S_{u,v,0}] + [l_{u,v,1} \cdot 100 + (1000 - 100) \cdot S_{u,v,1}]) = 0
\end{align*}

\noindent
The statement is true for $l_{u,v} = [0, 1]$ and $S_{u,v} = [0, 0.\bar{44}]$. 
The delay on the link is then calculated as follows:

\begin{align*}
\LD(u,v) =& \sum^{|\DCU_{u,v}|-1}_{p=0} [l_{u,v,p} \cdot \DCD_{u,v,p} + (\DCD_{u,v,p+1} - \DCD_{u,v,p}) \cdot S_{u,v,p}] \\
  =& [l_{u,v,0} \cdot 10 + (10 - 10) \cdot 0] + [1 \cdot 10 + (80 - 10) \cdot 0.\bar{44}] \approx \unit[41]{ms}
\end{align*}

\noindent
The end-to-end delay of an application is then the sum of delays on the links traversed by the application.
We denote the end-to-end delay of application $a$ with $\AD(a)$:

\begin{equation}
%\label{eq:constr_adelay}
\AD(a) := \sum_{(u,v) \in \mathcal{E}} \LD(u,v) \cdot \Fx_{a,u,v}
\end{equation}
\vspace{0.1em}

%\noindent
%As $\LD(u,v) \cdot \F_{a,(u,v)}$ is quadratic, we re-write it to a linear constraint.
%We introduce the function $\FLD_{a, (u,v)}$, which describes the delay of an application flow $a$ on a single link $(u,v)$.
%Furthermore we set $D^{(\text{max})} := 100000$ to a arbitrary large number as the maximal observable delay.
%$\FLD_a$, $\forall a \in \Ax$, is subject to:
%
%\vspace{-0.4em}
%\begin{equation}
%%\label{eq:constr_adelay}
%0 \leq \FLD_{a, (u,v)}
%\end{equation}
%
%\begin{equation}
%%\label{eq:constr_adelay}
%\FLD_a(u,v) \leq D^{\text{max}} \cdot \F_{a, (u,v)}
%\end{equation}
%
%\begin{equation}
%%\label{eq:constr_adelay}
%0 \leq \LD(u,v) - \FLD_{a, (u,v)}
%\end{equation}
%
%\begin{equation}
%%\label{eq:constr_adelay}
%\LD(u,v) - \FLD_{a, (u,v)} \leq \D^{\text{max}} \cdot (1 - F_{a, (u,v)})
%\end{equation}
%\vspace{-0.5em}
%
%\noindent
%The end-to-end delay of application $a$ can then be defined as $\AD_a$:
%
%\begin{equation}
%%\label{eq:constr_adelay}
%\AD(a) := \sum_{(u,v) \in \mathcal{E}} \FLD_a(u,v)
%\end{equation}
%\vspace{0.1em}

\noindent
Finally, the delay of the flow is not allowed to exceed the requirement:

\vspace{-0.4em}
\begin{equation}
\label{eq:last_delay_constr}
\AD(a) \leq \Dx(a) \quad \forall a \in \Ax
\end{equation}

\subsection{Problem Complexity and Possible Solving Strategies}

The optimization formulation combines variations of the non-splittable multi-commodity flow problem (routing) \cite{hu1963multi} and of the knapsack problem (balancing demand and utility), both known to be NP-hard.
Hence, approximation algorithms have to be found to solve the formulation in a reasonable runtime for larger topologies with potentially multiple bottleneck links and a large number of simultaneous applications.
The efficient and fast solving of the problem is out of scope of this work and is left to future work. 
This work provides the necessary abstractions and implementation proof that once the allocation decision is made, it can be efficiently and accurately be implemented in the network. 
As with other network resource allocation problems, such as the virtual network embedding (VNE) problem, the efficient solving of the theoretical problem can now be explored independently of the implementation concepts.

\begin{figure}%[]
\centering
\subfigure[Solving Time]{\includegraphics[width=95pt]{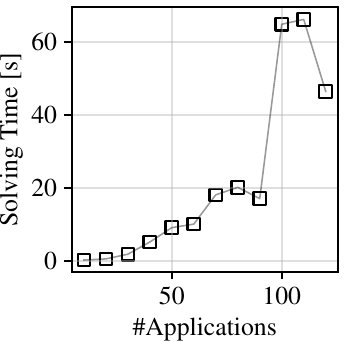}\label{fig:31_solving_time}} \hspace{20pt}
\subfigure[Variables]{\includegraphics[width=101pt]{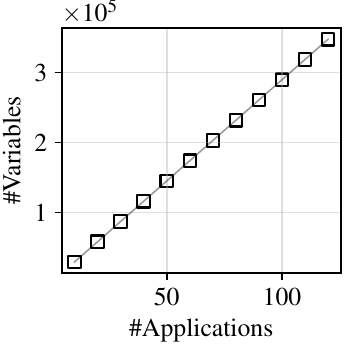}\label{fig:31_variables}} \hspace{15pt}
\subfigure[Constraints]{\includegraphics[width=101pt]{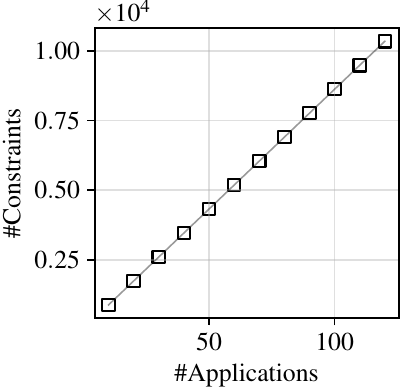}\label{fig:31_constraints}}
\vspace{-10pt}
\caption{
Problem size and solving time of the optimization formulation for increasing number of applications ($|\Ax|$) sharing \textit{one} bottleneck link. 
Maximum of \unit[66.2]{s} solving time for 110 applications.
2896 variables and 86 constraints for each additional application.
}
\label{fig:31_solving_time_all}
\Description[Linear increase of problem size.]{Problem size increases linearly with increasing number of applications.}
\vspace{-2pt}
\end{figure}

Solving the problem for our evaluation scenario (one bottleneck link, $\leq 120$ applications) takes on average less than one minute on a standard eight-core Intel Core i7-4770 \unit[3.4]{GHz} desktop PC with \unit[32]{GB} RAM using the commercial Gurobi\footnote{http://www.gurobi.com/} solver.
In detail, Figures~\ref{fig:31_solving_time} to \ref{fig:31_constraints} illustrate the solving time, total number of variables and total number of constraints of the problem instances with increasing number of applications with one bottleneck link.
The solving time stays below \unit[10]{s} up to approximately 50 applications. 
Above 90 applications the solving time increases drastically up to \unit[66.2]{s}. 
Afterwards, when a high number of applications does not leave much room for allocating higher utility values, the solving time decreases again.
Figures~\ref{fig:31_variables} and \ref{fig:31_constraints} show that the number of variables increases linearly with the number of applications with 2889 variables and 86 constraints for each additional application.
Thus, the total number of variables and constraints depends on the number of applications, on the used quantification of the utility and link delay functions and on the size of the network topology.

One greedy algorithm for finding a viable solution could be to start with a target utility of 1.0 for all application flows and shortest path routing.
Subsequently the utility can be increased by increments of 0.1 in a round-robin order until an allocation is reached where no application's utility can be increased anymore without violating capacity or delay constraints.
One problem with this algorithm is that it does not find sophisticated solutions where the utilization of one path is kept low to support low volume-low delay applications, e.g., web browsing, and other paths are dedicated to batch transfers, e.g., file download.

Despite sophisticated approximations there may be delay between a change to the global state, e.g., a new application, and the availability of a new allocation. 
Applications may have to wait before they can join the network, lower priority applications have to be disconnected or some throughput has to be reserved for yet unknown applications. 
This reserved capacity can then be allocated to new applications without requiring the solver to recalculate.

\section{Experiment Value Setup}%: Two Applications Sharing One Link}
\label{appendix:valuesetup}

In this section we describe the value setup used in the experiments in this paper for a two applications setup.
For increasing number of applications, all variables with dependency on $\Ax$ increase in size along their first dimension.
Table~\ref{tab:exp_value_setup} summarizes the problem input variables.
In the experiments we have a network topology $G$ with one bidirectional link ($\mathcal{E}=[(0,1),(1,0)]$) between two nodes ($\mathcal{V} := [0,1]$).
The link is shared by two application flows ($\Ax := [0,1]$) which both send data from node 0 to node 1 ($\Sx := [0,0]$, $\Tx := [1,1]$).
The link has a capacity of \unit[100]{Mbps} in both flow directions ($\Cx := [100~Mbps,100~Mbps]$). 
The link delay is modeled as constant with \unit[2]{ms} for our managed scenarios where combined paced throughput does not exceed the link capacity ($\DCU := [0, 100~Mbps]$, $\DCD := [2~ms,2~ms]$).
$\UTP$, $\UD$ and $\Ux$ describe the quantized utility functions from Fig.~\ref{fig:util_grids}.
An example for the quantization of the utility functions can be found in Section \ref{subsec:appendix_utilselection}.

\begin{table}[b]
%\vspace{2em}
	\caption{Experiment Value Setup For Two Applications}
	\vspace{-10pt}
    \begin{tabular}{ll}
%    \hline
    Symbol and value & Description \\ \hline %\hline
    \multicolumn{2}{|c|}{Problem Input Variables} \\ \hline
    $G(\mathcal{V} := [0,1],\mathcal{E}=[(0,1),(1,0)])$ & Network topology with two nodes and unidirectional links between them. \\ 
    $\Ax := [0,1]$ & 2 applications competing for the link. \\
    $\Sx := [0,0]$, $\Tx := [1,1]$ & Application flows are from node 0 to node 1. \\
    $\DCU := [0, 100~Mbps]$, $\DCD := [2~ms,2~ms]$ & Constant delay of \unit[2]{ms} for link (0,1). \\
    $\Cx := [100~Mbps,100~Mbps]$ & The link has speed of 100 Mbps in both directions. \\ 
    $\UTP$, $\UD$, $\Ux$ & Quantized utility functions. See \ref{subsec:appendix_utilselection} for an example. \\
    \end{tabular}
\label{tab:exp_value_setup}
\end{table}